# Competing 4f-electron dynamics in Ce(Ru$_{1-x}$Fe$_x$)$_2$Al$_{10}$ (0≤x≤1.0): magnetic ordering emerging from the Kondo semiconducting state

D.T. Adroja[1,$], A.D. Hillier[1], Y. Muro[2], J. Kajino[3], T. Takabatake[3], P. Peratheepan[4], A.M. Strydom[4], P.P. Deen[5], F. Demmel[1], J.R. Stewart[1], J.W. Taylor[1], R.I. Smith[1], S. Ramos[6] and M. A. Adams[1]

[1] ISIS Facility, Rutherford Appleton Laboratory, Chilton, Didcot Oxon, OX11 0QX, UK
[2] Liberal Arts and Sciences, Faculty of Engineering, Toyama Prefectural University, Imizu 939-0398, Japan
[3] Department of Quantum matter, ADSM, and IAMR, Hiroshima University, Higashi-Hiroshima, 739-8530, Japan
[4] Physics Department, University of Johannesburg, PO Box 524, Auckland Park 2006, South Africa
[5] European Spallation Source, St Algatan 4, Box 176 Lund Sweden 221 00
[6] Diamond Light Source, Harwell Science & Innovation Campus, Didcot, Oxon OX11 0DE

(Dated: 19$^{th}$ May 2013)

We have carried out muon spin relaxation (μSR), neutron diffraction and inelastic neutron scattering (INS) investigations on polycrystalline samples of Ce(Ru$_{1-x}$Fe$_x$)$_2$Al$_{10}$ (x=0, 0.3, 0.5, 0.8 and 1) to investigate the nature of the ground state (magnetic ordered versus paramagnetic) and the origin of the spin gap formation as evident from the bulk measurements in the end members. Our zero-field μSR spectra clearly reveal coherent two-frequency oscillations at low temperature in x=0, 0.3 and 0.5 samples, which confirms the long-range magnetic ordering of the Ce-moment with T$_N$=27, 26 and 21 K respectively. On the other hand the μSR spectra of x=0.8 and x=1 down to 1.4 K and 0.045 K, respectively exhibit a temperature independent Kubo-Toyabe term confirming a paramagnetic ground state. The long-range magnetic ordering in x=0.5 below 21 K has been confirmed through the neutron diffraction study. INS measurements of x=0 clearly reveal the presence of a sharp inelastic excitation near 8 meV between 5 K and 26 K, due to an opening of a gap in the spin excitation spectrum, which transforms into a broad response at and above 30 K. Interestingly, at 4.5 K the spin gap excitation broadens in x=0.3 and exhibits two clear peaks at 8.4(3) and 12.0(5) meV in x=0.5. In the x=0.8 sample, which remains paramagnetic down to 1.2 K, there is a clear signature of a spin gap of 10-12 meV at 7 K, with a strong Q-dependent intensity. Evidence of a spin gap of 12.5(5) meV has also been found in x=1. The observation of a spin gap in the paramagnetic samples (x=0.8 and 1) is an interesting finding in this study and it challenges our understanding of the origin of the semiconducting gap in $CeT_2Al_{10}$ (T=Ru and Os) compounds in terms of hybridization gap opening only a small part of the Fermi surface, gapped spin waves, or a spin-dimer gap.

# I. Introduction

Ce and Yb-based compounds exhibit a rich variety of novel phenomena, such as heavy electron behavior, mixed valence behavior, reduced magnetic moment ordering, Kondo insulator or Kondo semiconductor, spin and charge gap formation, charge and spin density waves, metal-insulator transition, unconventional superconductivity, spin-dimer formation, non-Fermi-liquid (NFL) behavior and quantum criticality [1-10]. These phenomena arise due to the presence of strong hybridization between localised 4f-electrons and conduction electrons [4, 9]. Recently the $CeT_2Al_{10}$ (T=Fe, Ru and Os) compounds have attracted interest in condensed matter physics, both experimentally and theoretically, due to the remarkable physical properties they exhibit [11-22]. For example the opening of a spin and charge gap, anisotropic hybridization and charge density modulation have been suggested [11-12, 17-20]. The Ru and Os compounds order antiferromagnetically at $T_N$=27 and 29 K, respectively, while the Fe compound remains paramagnetic down to 50 mK [11-13]. The magnetic susceptibility shows that $CeFe_2Al_{10}$ is a valence fluctuation system with strong anisotropic hybridization [13], while $CeRu_2Al_{10}$ shows the $Ce^{3+}$ ionic state, but $CeOs_2Al_{10}$ shows a strong hybridization effect [12]. Furthermore, $CeFe_2Al_{10}$ exhibits Kondo semiconducting behavior with a transport gap of 15 K, while NMR and heat capacity studies reveal a larger value of the gap, 125 K and 100 K, respectively [13, 21, 22]. The Kondo semiconductor behavior observed in $CeFe_2Al_{10}$ bears similarity with that of the well-known Kondo semiconductors CeNiSn and CeRhSb [23, 24]. Therefore systematic investigations of $CeT_2Al_{10}$ (T = Fe, Ru, and Os) with different values of the Kondo temperature $T_K$ (or hybridization) are necessary to reveal the role of the 4f-electrons and conduction electrons hybridization in the mysterious phase transition and gap formation. The Ce$T_2$Al$_{10}$ series of compounds offers therefore an isoelectronic platform from which to study a systematic increase in the electronic hybridization

Recently, µSR and neutron scattering studies on CeT$_2$Al$_{10}$ (T=Ru and Os) have been performed [25-30]. The µSR studies in CeT$_2$Al$_{10}$ (T=Ru and Os) revealed the presence of small internal fields 20-150G (depending on the muon sites) [25-27], respectively at the muon stopping site in zero-field indicating unambiguously long-range magnetic ordering of the Ce$^{3+}$ moment in both compounds. Surprisingly, inelastic neutron scattering study (INS) clearly indicated a spin-gap formation of 8 meV and 11 meV in T=Ru and Os, respectively, in the ordered states [27, 28]. The gap is nearly temperature independent very close to T$_N$ in both compounds, but then abruptly develops into a broad quasi-elastic/inelastic response above T$_N$ [27, 28]. By raising the temperature still further (above 40 K), the INS response becomes very broad, with quasi-elastic character in both compounds [27, 28]. The observation of a spin gap in these compounds is in good agreement with predictions based on a theoretical model for a spin-dimer formation pertinent to this class of compounds which has recently been put forward by Hanzawa [31, 32]. However, our recent spin wave studies on single crystalline samples of CeT$_2$Al$_{10}$ (T=Ru and Os) [33] and also those by Robert et al on T=Ru [34] reveal the gapped spin wave excitations (gap ~ 4-5 meV at AFM zone centre).

On the alloy system Ce(Ru$_{1-x}$Fe$_x$)$_2$Al$_{10}$ (x=0 to 1), magnetic and thermal measurements have revealed that T$_N$ remains nearly constant up to x=0.7 and then abruptly disappears at x=0.8 (no long range ordering down to 1.2 K) [35]. Therefore this system provides an ideal choice to tune the strength of hybridization across the series, as the Ce ions in x=0 are close to 3+ state and those in x=1 are in mixed valence (or valence fluctuating state). We therefore have carried out μSR, neutron diffraction, inelastic neutron scattering and x-ray absorption near-edge structure (XANES) studies on Ce(Ru$_{1-x}$Fe$_x$)$_2$Al$_{10}$ to shed light on the nature of the spin gap formation and the ground state of the Ce ion in this system. Considering that the ordered state moment of the Ce ion is very small, 0.34(2)μ$_B$ in CeRu$_2$Al$_{10}$ [25], μSR as an exceptionally sensitive microscopic probe is ideally suited to this problem. Inelastic neutron scattering gives direct information about the magnitude of the spin-gap energy and its temperature and wave-vector (Q) dependency, which are important to understand the nature of the mechanism of the spin gap formation [4, 36, 37].

## II. Experimental details

The polycrystalline samples of Ce(Ru$_{1-x}$Fe$_x$)$_2$Al$_{10}$ (x=0, 0.3, 0.5, 0.8 and 1) and non-magnetic phonon reference compounds LaRu$_2$Al$_{10}$ and LaFe$_2$Al$_{10}$ were prepared by argon arc melting of the stoichiometric constituents with the starting elements , Ce/La 99.9% in purity, Ru and Fe 99.9% and Al 99.9999%. The samples were annealed at 800 $^o$C for one week in an evacuated quartz ampoule. Phase characterization using neutron powder diffraction proved the samples to be practically single-phase. The impurity phase amounted to about 3 volume-% in $x=0.5$, but its chemical composition is not known at present.

For the zero-field (ZF) µSR experiments, the powdered samples (thickness ~1.5mm) were mounted onto a 99.995+% pure silver plate using GE-varnish and were covered with 18 micron silver foil. We used the MuSR spectrometer in longitudinal geometry at the ISIS Pulsed Neutron and Muon Source, UK. At the ISIS facility, a pulse of muons is produced every 20 ms and has a full-width at half-maximum (FWHM) of ~70 ns. These muons are implanted into the sample and decay with a half-life of 2.2 µs into a positron which is emitted preferentially in the direction of the muon spin axis. These positrons are detected and time stamped in the detectors which are positioned before, F, and after, B, the sample. From the measured positron counts in the F and B detectors, $N_F(t)$ and $N_B(t)$, respectively, the asymmetry of the muon decay, $G_z(t)$ is determined using

$$G_z(t)=(N_F(t)-\alpha N_B(t))/(N_F(t)+ \alpha N_B(t)) \qquad (1)$$

where $\alpha$ is a calibration coefficient [27].

The neutron diffraction measurements at 300 K were performed using the GEM diffractometer at ISIS Facility. The low temperature neutron powder diffraction measurements on $x = 0.5$ sample were carried out using the OSIRIS spectrometer in diffraction mode. The sample was mounted in a 20 mm diameter Al-can, which was cooled down to 5 K using a standard top-loading closed cycle refrigerator (TCCR) with He-exchange gas around the sample for thermalization. The inelastic neutron scattering measurements on x = 0, 0.3 and 0.5 were carried out using the MARI time-of-flight (TOF) chopper spectrometer and on x=0.8 and 1 were carried out using the high neutron flux MERLIN TOF spectrometer at ISIS Facility. The powder samples (mass ~20g) were wrapped in a thin Al-foil and mounted inside a thin-walled cylindrical Al-can, which was cooled down to 4.5 K inside a TCCR-with He-exchange gas around the samples. The measurements were performed with various selected incident neutron energies ($E_i$) between 20 meV and 100 meV.

The Ce $L_3$-edge x-ray absorption near-edge structure (XANES) of x=0 and 1 compounds and the reference compound $CeCoSi_3$ was measured in transmission mode (at 300 K) using beamline B18, the Core EXAFS (Extended X-ray Absorption Fine Structure) Beamline, at the Diamond Light Source, UK. Samples were prepared by grinding the polycrystalline material into a fine powder, mixing it with cellulose and pressing the mixture into pellets.

## III. Results and discussions

### (1) Structural study using powder neutron diffraction

Figure 1 shows the neutron diffraction patterns of Ce(Ru$_{1-x}$Fe$_x$)$_2$Al$_{10}$ (x=0, 0.3, 0.5, 0.8 and 1) at 300 K collected in the 2θ ~ 60 degrees detector banks of GEM. In order to investigate the change in the lattice parameters, unit cell volume and Ce-X (X=Ce, Al, and Ru/Fe) distances with Fe composition (x), we have carried out a full structural refinement using the GSAS program. Details of the structural model used in the present analysis is given in refs. [25, 27]. The refinement confirms that the compounds crystallize in the orthorhombic YbFe$_2$Al$_{10}$-type structure (space group Cmcm, No. 63). In this caged-type structure the Ce atom is surrounded by a polyhedron formed by 4 Ru/Fe and 16 Al atoms and forms a zigzag chain along the orthorhombic c-axis [15]. The refined lattice parameters, unit cell volume and the selected Ce-Ce, Ce-Al and Ce-Ru(Fe) interatomic distances are plotted in Figs. 2 and 3, respectively. One can see from Fig.2 that the lattice parameters (*a, b, c*) decrease gradually with increasing Fe composition (x). The lattice parameters *b* and *c* and unit cell volume of x=0.8 and 1 show weak deviation from the linearity, which we attribute due to the increase in the mixed valence nature of the Ce with x and especially for x=1. The change in the unit cell volume is about 3.5% while going from Ru to Fe. Furthermore, the nearest neighbour Ce-Ce and Ce-Ru(Fe) distances also decrease linearly with increasing x. On the other hand although Ce-Al$_i$ (i=2 and 5) distances decrease linearly with x, Ce-Al$_i$ (i=1, 3 (especially d2 of i=3) and 4) distances reveals some non-linearity with x. Further the Ce-Al$_3$ (d2) distance exhibits a noticeable slope change above and below x=0.5, suggesting the change in the hybridization between the Ce 4f and Al$_3$ 3p electrons. This hybridization in CeRu$_2$Al$_{10}$ may stabilize the wave function of the ground state doublet which has a butterfly shape elongating along the Al$_3$ atom [12, 38].

**(2) μSR measurements**

Figure 4 (a-h) shows the zero-field (ZF) μSR spectra at various temperatures of Ce(Ru$_{1-x}$Fe$_x$)$_2$Al$_{10}$ (x=0, 0.3, 0.5, 0.8 and 1). For comparison purposes we refer to the data of x=0 from ref [25]. At 35 (or 30) K we observe a strong damping at shorter relaxation time (Fig.4d-h), and the recovery at longer times, which is a typical muon response to nuclear moments, described by the Kubo-Toyabe formalism [39], arising from a static distribution of the nuclear dipole moments. Here it arises from the $^{101}$Ru (I=5/2) and $^{27}$Al (I=5/2) nuclear moment contributions (I=0 for Ce and $^{56}$Fe, i.e. zero nuclear contribution). Above the anomaly at 28 K, i.e. in the paramagnetic state, the μSR spectra can all be described by the following equation (see Figs. 4d-h):

$$G_z(t) = A_0 \left( \frac{1}{3} + \frac{2}{3} (1 - (\sigma t)^2) \exp\left(-\frac{(\sigma t)^2}{2}\right) \right) \exp(-\lambda t) + C \qquad (2)$$

where $A_0$ is the initial asymmetry, σ is nuclear depolarization rate, $\sigma/\gamma_\mu = \Delta$ is the local Gaussian field distribution width, $\gamma_\mu$=13.55 MHz/T is the gyromagnetic ratio of the muon, λ is the electronic relaxation rate and C is a constant background. It is assumed that the electronic moments give an entirely independent muon spin relaxation channel in real time. The value of σ was found to be 0.32-0.36 μs$^{-1}$ (depending on x) from fitting the spectra of 35/30 K to Eq.(2) and was found to be temperature independent above 35 K. It is to be noted that using a similar value of the σ Kambe et al [26] have suggested 4$a$ as the muon stopping site in $CeRu_2Al_{10}$, while for $CeOs_2Al_{10}$ the muon stopping site was assigned to the (0.5, 0, 0.25) position [27].

It is interesting to see a dramatic change in the time-evolution of the μSR spectra with temperature for x=0, 0.3 and 0.5 (Figs.4a-c), while that of x=0.8 and 1 do not show any noticeable change with temperature (Figs.4d & h). The spectra below 27 K are best described by two oscillatory terms and an exponential decay, as given by the following equation

$$G_z(t) = \left(\sum_{i=1}^{2} A_i cos(\omega_i t + \varphi) \exp\left(-\frac{(\sigma_i t)^2}{2}\right)\right) + A_3 \exp(-\lambda t) + C \qquad (3)$$

where $\omega_i = \gamma_\mu H^i_{int}$ is the muon precession frequencies ($H^i_{int}$ is the internal field at the muon site), $\sigma_i$ is the muon depolarization rate (arising from the distribution of the internal field) and φ is the phase.

In Fig. 5 (a-c) we have plotted the muon precession frequencies (or internal fields) at the muon sites as a function of temperature for x=0, 0.3 and 0.5. This shows that the internal fields appear just below 27 K for x=0, showing a clear evidence for long-range magnetic ordering. A very similar presence of internal fields has been observed below 26 K in x=0.3 and below 22 K in x=0.5 indicating the presence of long-range magnetic ordering. Further it is very important to mention that the asymmetry $A_3$ drops nearly 2/3 and the relaxation rate exhibits small drops at $T_N$ for x=0, 0.3 and 0.5 (figure not shown), which confirms that the magnetic ordering is observed in the full volume of the samples and hence is bulk in nature. The value of $T_N$ estimated from µSR study is plotted as a function of x in Fig.2 using open (red) diamond symbols. It is interesting to note that the observed two muon precession frequencies are about a factor 5 different in x=0 across the entire $T<T_N$ temperature range, while the difference is found to decrease gradually and reaches a factor of 1.6 (at the lowest temperature) for x=0.5. The low-temperature upturn in precession frequencies appears to be a feature characteristic only of the Fe-containing compounds. Furthermore the values of the highest frequencies decrease with increasing x (going from x=0 to 0.5), which may indicate that the ordered state Ce moment is reducing with x. This is in agreement with the observed susceptibility behavior [35], which indicates that with increasing x the hybridization increases and the valence of the Ce ion shifts toward mixed valence (or 4+) value. The small value of the frequencies/internal fields observed in x=0 to 0.5 are in agreement with the small ordered state magnetic moment of the $Ce^{3+}$ ion observed through the neutron diffraction for x=0 [25] and x=0.5 discussed in the next section.

Now examining the temperature dependence of the frequencies, we can see that there is a dip in the frequency (see Fig. 5a), which occurs around 13 K for x=0. In contrast, a rise in the frequency below 10 K (Fig.5b) and 5 K (Fig.5c) for x=0.3 and 0.5 respectively is observed. The occurrence of the dip in x=0, which has also been observed in the µSR study of $CeOs_2Al_{10}$ at 10-15 K [27], may have some relation with a super lattice formation observed in the recent electron diffraction study of $CeOs_2Al_{10}$ [12]. Moreover, below $T_N$ the first component of the depolarisation rates for x=0 exhibits a strong temperature dependence (Fig. 5d-f, right) and a weak anomaly with decreasing temperature, while the second component of the depolarisation rates is weakly temperature dependent and exhibits a sharp rise below 5 K. In principle this could originate from various phenomena related to a change in the distribution of internal fields associated with a small change in the moment values or modulation. The support for this argument comes from our preliminary neutron diffraction study at high d-spacing (up to 40 Å) on $CeRu_2Al_{10}$, which reveals the presence of a weak and broad peak near d=32 Å that exists only below 5 K [29]. Further, for x=0.3 and 0.5 the observed anisotropy of the depolarization rates (observed in x=0) becomes smaller with increasing x.

### (3) Magnetic neutron diffraction study on x=0.5

In order to investigate the magnetic structure of the x=0.5 compound, we have carried out a neutron diffraction study of x=0.5 between 5 and 35 K (Figs. 6a & b). Comparing the data collected at 5 K and 35 K, we observe two additional reflections (and one weak reflection on the top of nuclear peak) at 5 K. Further the background at 5 K is reduced compared to that at 35 K (the data are scaled to match the background), which indicates that the observed additional reflections are magnetic in nature. We can index the observed magnetic reflections using the same propagation vector $k$ =(1 0 0) that was used for the parent compound $CeRu_2Al_{10}$ [30]. It should be noted that for $CeOs_2Al_{10}$ Kato et al used $k$ =(0 1 0), [25], which is related to [1 0 0] by the reciprocal lattice vector (-1 1 0). Further the absence of [0 0 $\ell$]-type magnetic Bragg peaks indicates that the moment is along the c-axis as observed in $CeRu_2Al_{10}$. In order to estimate the size of the moment we have carried out a simulation of the 5 K data, using the same magnetic structure proposed for $CeRu_2Al_{10}$. Our simulation gave an estimate on the ordered state Ce moment ~0.17(4) $\mu_B$ for x=0.5 (Figs.6 c-d) compared with 0.34(2) $\mu_B$ for x=0 [25]. This reduction of the moment in x=0.5 is expected due to the presence of strong hybridization between 4f- and conduction electrons and in agreement with the observation of very small internal magnetic fields seen in the μSR data as discussed above. In addition, in the *x*=0.5 compound it is anticipated that the Kondo semiconducting state which is an extreme case of 4f and conduction electrons hybridization may already be in evidence. In order to investigate the temperature dependent order parameters, we performed a diffraction study (at a selected d-range) for various temperatures between 5 K and 35 K. Figs. 7a & b show the integrated intensity of [1 0 1] magnetic Bragg peak and background, respectively. It is clear that below 22 K we have long-range magnetic ordering. Further, the temperature dependent intensity first increases linearly below 22 K and then saturates below 15 K. The observed rise below $T_N$ is slightly weaker than that observed in $CeRu_2Al_{10}$ [25] and $CeOs_2Al_{10}$ [30], which might be due to the effect of Ru/Fe disorder on the exchange parameters.

**(4) Inelastic neutron scattering study**

The compound $CeRu_2Al_{10}$ has a spin gap of 8 meV at temperatures below 29 K [11-12, 20, 28]. Our μSR spectra of $Ce(Ru_{1-x}Fe_x)_2Al_{10}$ changed dramatically between x=0.5 and x= 0.8. Therefore, it is of interest to investigate the composition dependence of the spin gap value and its Q- and temperature dependence across the series using inelastic neutron scattering. It is to be noted that initial INS measurements on $CeRu_2Al_{10}$ were carried out using a triple axis spectrometer (TAS) [28], which provided only limited Q-information compared to the present TOF study that allows surveying a larger volume of Q-E space in one measurement and hence provides a wealth of information. The TOF studies are important for the present systems, as we need to untangle two contributions, spin wave versus hybridization gap. Therefore, we report the compositions and temperature dependent INS spectra of $Ce(Ru_{1-x}Fe_x)_2Al_{10}$ (x=0, 0.3, 0.5, 0.8 and 1) in this section. We have also measured the non magnetic phonon reference compounds $LaRu_2Al_{10}$ and $LaFe_2Al_{10}$. A detailed report on the inelastic neutron scattering investigations on $CeOs_2Al_{10}$ compound can be found in ref. [27, 40]

*4.1 Spin gap in the magnetic ordered state (x=0-0.5)*

Figure 8 displays the color-coded plot of the scattering intensity, energy transfer versus momentum transfer, of x=0, 0.3 and 0.5 along with the reference compound $LaFe_2Al_{10}$ measured at 4.5 K on the MARI spectrometer. The data of $LaRu_2Al_{10}$ were used to subtract phonon contribution in the samples with low Fe content, i.e. x=0 and 0.3. The phonon contribution was subtracted by scaling the La-data by the cross section ratio of the Ce-compounds and La-compounds and then subtracting from the Ce-compounds data (we called this method-1), for detail see Ref. [41]. There is a clear magnetic excitation centred around 8 meV in x=0, which was found in the TAS study by Robert et al [28]. The value of the peak position can be taken as a measure of the spin gap energy in these compounds [37]. The spin gap energy of 8 meV is in good agreement with the value determined from the exponential behavior of the observed magnetic susceptibility, specific heat and NMR studies [11, 21, 22]. Further in x=0.3 and 0.5 the magnetic scattering broadens and the intensity is considerably reduced compared to x=0. To see the linewidth ($\Gamma$) and intensity clearly we have plotted the data in 1D (Q-integrated between 0 and 2.5Å) energy cuts (see Fig. 9) taken from the 2D colour plots. From Fig. 9 it is clear that we have spin gap type excitations in all three compounds. The presence of a spin gap in the excitation spectrum in x=0.5 (and also in x=1) was also supported through the low energy and high resolution ($\Delta E = 25\mu eV$ at elastic line) INS measurements on OSIRIS (data not shown here), which did not reveal any clear sign of a quasi-elastic scattering below 2 meV at 5 K.

In the following section, we discuss the temperature dependence of the spin gap excitation in x=0, 0.3 and 0.5. Fig. 10 shows the estimated magnetic scattering at various temperatures for x=0 (left), 0.3 (middle) and 0.5 (right). It is to be noted that as we did not measure La(Ru$_{1-x}$Fe$_x$)$_2$Al$_{10}$ with x=0.5, we used the ratio of the high-Q and low-Q data of LaFe$_2$Al$_{10}$ (i.e. Ratio = [S(High-Q, ω)/S(Low-Q, ω)]$_{La}$) to estimate the magnetic scattering in x=0.5: $S_M(Q,\omega)$=S(Low-Q, ω)$_{Ce}$- S(High-Q, ω)$_{Ce}$/Ratio (we call this as method-2). It is clear that in x=0 the magnetic scattering remains practically temperature independent up to 20 K and then decreases abruptly with increasing temperature (Fig.10 left). At 40 K the scattering becomes quasi-elastic. A very similar behavior has been observed in x=0.3 and 0.5. In order to investigate the involvement of prevailing inelastic type energy excitations, we have analysed the temperature dependent magnetic scattering (S (Q ω)) using a Lorentzian lineshape [27] and fits are shown in Fig. 10. The scattering law, S (Q ω), is related to the imaginary part of the dynamical susceptibility: S (Q ω)=1/(1-exp(-ℏω/K$_B$T))*Im χ, the symbols have their usual meaning. Further Im χ/ω can be taken as a Lorentzian form [27]. It is to be noted that the optical study reveals the presence of charged density wave (CDW) gap above T$_N$ in both CeRu$_2$Al$_{10}$ and CeOs$_2$Al$_{10}$ [18, 19]. The solid line shows the fit using an inelastic peak (we allowed the peak position to vary) and the dotted line shows the fit using a quasi-elastic peak position (i.e. peak position was fixed at zero energy). The data of x=0.5 show two INS excitations and the origin of this is discussed further below.

Figure 11 shows the temperature dependent parameters estimated from the fit to the data for x=0 and 0.3 (filled circles are INS fits and open circles are quasi-elastic fits). Figs.11a&d show the estimated magnetic susceptibility for both compounds. Thereby, we assumed that van Vleck contribution from the high energy CEF is small at low temperature. For x=0 the estimated susceptibility is close to that measured using a SQUID magnetometer shown by the small blue filled circles. This is also the case for x=0.3 and 0.5, when compared with the reported single crystal susceptibility [35]. Figs.11b&e show the temperature dependent linewidth, $\Gamma(T)$, and Figs.11c&f show the temperature dependent peak position, $\Delta(T)$, (i.e. spin gap). For comparison purpose, we have also plotted the data of x=0 from Robert et al [28] using open squares. It is clear that $\Gamma(T)$ of x=0 decreases below $T_N$. We have analysed $\Gamma(T)$ using two models: (1) $\Gamma(T) \sim T^2$ and (2) exponential behavior, $\Gamma(T) \sim e^{(-\Delta(0)/k_B T)}$, where $k_B$ is Boltzmann's constant. The exponential relation for $\Gamma(T)$ is found to describe the data much more reliably with gap value of $\Delta(0) \sim 8.0(2)$ meV, which is in good agreement with the peak position observed at 4.5 K. Now we compare the magnitude and temperature dependent spin gap of x=0 to that predicted by Hanzawa's theoretical model based on the nearest neighbour (NN) Ce-Ce RKKY interactions in mean field that predicts a spin-spiral gap and its temperature dependence [31]. The dotted line (in the bottom of Fig.11c) shows the calculation from Hanzawa's model (without any scaling factor) [31]. The temperature dependence of the observed spin gap and theoretically predicted spin gap is similar in behavior just below $T_N$, but there is clear evidence in the experimental data to support the existence of the spin gap just above $T_N$ (possibly upto 33 K) in x=0 and also in x=0.3 (up to 35K). The optical study on $CeRu_2Al_{10}$ also shows the existence of a gap above $T_N$ through the effective electron number $N_{eff}$, which is related to the gap $N_{eff} \sim \Delta^2_{opt}$ [19]. For comparison we have also plotted $N_{eff}^{1/2}$ (open triangles, normalized to INS gap at the lowest temperature) in Fig. 11c [19]. A very similar situation has also been observed for $CeOs_2Al_{10}$ through an optical study [18], where a CDW gap (or $\Delta_{opt}$) exists up to 39 K, and also from our recent INS study [40], where we have seen an INS peak surviving up to 38 K. The INS data of x=0.5 (Fig.10 right) also reveal the possibility of the spin gap (solid line) at 35 K. If we take the value of the quasi-elastic linewidth as a measure of Kondo temperature $T_K$ (just above $T_N$, ideally one takes the value at T=0) then it shows that $T_K$ increases from 52(3) K in x=0, 83(5) K in x=0.3, to 110(10) K in x=0.5.

Now we discuss the Q-dependence of the energy integrated intensity between 7 and 10 meV at 4.5 K for x=0 (see Fig.1A(top) in the appendix), which is found to follow the $Ce^{3+}$ magnetic form factor squared ($F^2(Q)$), although some very weak oscillating feature around $F^2(Q)$ has been observed. The observed single ion type response could also be due to the fact that the first antiferromagnetic (AFM) Bragg peak (0 1 0) is situated at Q=0.61 Å$^{-1}$ (shown by a vertical arrow), from where the spin wave emerges, is very close to the edge of the low angle detectors' coverage and hence missing the full spin waves dispersion from (0 1 0). To investigate the conjecture of spin dimers forming in the magnetic ordered state, we have also analysed the data using an isolated dimer structure factor [42]. The red-dotted line in Fig.1A (bottom) shows the result of the fit, from which it is evident that this representation can hardly distinguish between a $F^2(Q)$ or a spin-dimer structure factor. Seeing that the spin gap in $CeRu_2Al_{10}$ and $CeOs_2Al_{10}$ opens in the magnetically ordered state, one would expect that the spin gap energy and its intensity would be strongly Q-dependent, especially from spin waves, which we have in fact observed in our single crystal study on $CeRu_2Al_{10}$ and $CeOs_2Al_{10}$ [33] and also by Robert et al in $CeRu_2Al_{10}$ [34].

Before discussing the spin gap in the paramagnetic compounds x=0.8 and 1, here we discuss the origin of the two peak-type structure observed in the Q-integrated intensity for x=0.5 at 4.5 K shown in Fig. 10 (right-hand panel). Fig.2A (in the appendix) shows the magnetic scattering measured with incident energy $E_i$=100 meV (top color plot). It is clear from this plot that we have observed dispersive excitations, which have strong intensity and an energy minimum near Q=0.5 Å$^{-1}$ and maximum energy near Q=1-1.5 Å$^{-1}$ (or above, see Fig.2A (bottom)). The presence of the dispersion suggests that the two peaks structure observed in x=0.5 for $E_i$=25 meV data is associated with the dispersive excitation in the powder sample and partly attributed to spin waves.

*4.2 Spin gap in the paramagnetic state (x=0.8)*

In order to elucidate the role of the hybridization or dimer gap formation in Ce(Ru$_{2-x}$Fe$_x$)$_2$Al$_{10}$, INS investigations are called for on the spin gap formation in the paramagnetic compounds x=0.8 and 1. Our µSR study discussed above confirms the paramagnetic ground state in these two compounds down to the lowest temperature (see Fig.4d&h). As we expected a very weak magnetic response in these compounds due to the presence of strong hybridization as evidenced through the magnetic susceptibility [35], we have investigated these compounds using the high flux MERLIN TOF spectrometer at ISIS. Figs. 12a&b show the color-coded plots of the scattering intensity for x=0.8 at 7 K and 94 K, respectively, along with the nonmagnetic phonon reference compound LaFe$_2$Al$_{10}$ at 7 K and 94 K (Figs. 12c&d). The x=0.8 compound at 7 K exhibits a clear sign of the spin gap-type magnetic scattering near 10 meV that is localised in Q (near 0.75 Å$^{-1}$) and transforms into a broad quasi-elastic response at 94 K. When we compared the data of LaFe$_2$Al$_{10}$ at 7 K, which does not show any sign of the scattering near 10 meV (at low-Q), with that of x=0.8 it is clear that the observed scattering near 10 meV in x=0.8 is due to the magnetic nature and possibly a spin gap formation. As the µSR study rules out the presence of long range magnetic ordering in this compound, the gap is not associated with spin waves in x=0.8. In Figs. 12e&f we have estimated the magnetic scattering in x=0.8 by subtracting the phonon scattering using LaFe$_2$Al$_{10}$ data. It is clear from these figures that the spin gap exists at 7 K, but is already collapsed at 94 K. We have also carried out INS measurements on MARI at 5 K, 35 K and 100 K with a selected incident energy of E$_i$=40 meV. Although the MARI data have comparably larger statistical deviations, it was clear that the 10 meV excitation does exist in x=0.8 at 5 K and 35 K, but at 100 K the response becomes quasi-elastic in agreement with the MERLIN data. This change in the response from a spin gap-type to a quasi-elastic line (Fig.13b) is in agreement with the observed broad maximum in the susceptibility at 50 K. This behavior has been observed in many spin gap systems, for example CeOs$_4$Sb$_{12}$ [36], CeRu$_4$Sb$_{12}$ [37, 41], CePd$_3$ [43] and CeFe$_4$Sb$_{12}$ [44]. A notable feature of the spin gap energies of these compounds measured through INS is their universal scaling relationship with the Kondo energy (T$_K$) derived from the maximum in the susceptibility [26, 37, 41, 44]. According to the single impurity model [37, 45], we can estimate the high temperature Kondo temperature T$_K$ through the maximum $T_{max}(\chi)$ in the bulk susceptibility as $T_K= 3*T_{max}(\chi)=150$ K (12.92 meV) for x=0.8. This shows that the spin gap of 10 meV observed through the INS study is in agreement with the scaling behavior. We would like to mention that the spin gap energy of $CeOs_2Al_{10}$ and $CeRu_2Al_{10}$ estimated from the INS measurements in the polycrystalline samples indeed followed this scaling behavior [27].

*4.2.1 Spin dimer versus anisotropic gap on the Fermi surface*

Now we discuss the Q-dependence of the spin gap intensity of x=0.8. In Fig.13c we have plotted the energy integrated (8-12meV) Q-dependent neutron scattering intensity from x=0.8 and $LaFe_2Al_{10}$ and in Fig.13d the estimated magnetic scattering. It is clear from this figure that the intensity of the 10 meV excitation in x=0.8 exhibits a clear peak near Q = 0.8 Å$^{-1}$ and does not follow $F^2(Q)$ behavior (typical for single ion type interaction) for $Ce^{3+}$. This behavior is different from that observed for many spin gap systems [37], which do not exhibit long range magnetic ordering. We also analysed the Q-dependent intensity using the isolated dimer structure factor in order to check the possibility of spin dimer formation as predicted by the Hanzawa model for $CeRu_2Al_{10}$ [31]. The fit to the dimer structure factor $I(Q) \sim Sin(Q\,d)/(Q\,d)$, where d is the Ce-Ce distance, is given by the red dotted line in Fig. 13d and fit gave d=5.07(4)Å, which is close to d=5.21(4)Å estimated through neutron diffraction study at 300 K. Although the peak intensity does not fit very well to the dimer structure factor, the peak positions are in agreement with dimer formation. Another possible interpretation of the observed spin gap in x=0.8 could be an anisotropic spin gap opening only on a small part of the Fermi surface or along a specific direction in Q-space. This is somewhat similar to the spin gap observed only along [0 0 l] direction in CeNiSn [46].  Additional support for an anisotropic spin gap may come from the anisotropic behavior of the pressure dependent resistivity of $CeRu_2Al_{10}$ below $T_N$,  which suggests that strong anisotropic gap is formed by a phase transition at 3GPa [16]. Considering the smaller unit cell volume of x=0.8 compared to x=0, which will act as a chemical pressure, the situation is similar between x=0.8 at ambient pressure and x=0 under pressure, and hence the gap could also be anisotropic in x=0.8. We have also measured x=0.8 with higher incident energy $E_i$=100 meV at 7 K (data not shown here). The estimated magnetic scattering showed one sharp inelastic peak near 10 meV, as in Fig. 13a and another broad ($\Gamma$~21 (2) meV) peak centred near 48 (1) meV.  Using these data along with $E_i$ = 40 meV data, we have estimated the total contribution to the susceptibility, $2.1(2) \times 10^{-3}$ (emu/mole), which is comparable to the single crystal susceptibility  for B//*a*, $4 \times 10^{-3}$ (emu/mole) [35]. Considering that the *a*-axis is the easy magnetization axis, the susceptibility values for B//*c* and B//*b* (not reported) will be ~$2 \times 10^{-3}$ and ~$1 \times 10^{-3}$ (emu/mole) (predicted using susceptibility value of $CeFe_2Al_{10}$ [13]) and hence the polycrystalline average will be close to $2.3 \times 10^{-3}$ (emu/mole), which is in good agreement with that estimated form our INS results. This confirms that INS study probes the bulk nature of the sample.

*4.3 Spin gap in the non-ordered Kondo insulating state (x=1)*

Finally we discuss our INS results of x=1 (i.e CeFe$_2$Al$_{10}$) measured on MERLIN with $E_i$=20 and 100 meV. We have also studied LaFe$_2$Al$_{10}$ with the same incident neutron energies to subtract the phonon background. It is clear from these data, as explained below, that we have a spin gap type magnetic scattering around 10-15 meV in CeFe$_2$Al$_{10}$, while weak phonon scattering in LaFe$_2$Al$_{10}$. The estimated magnetic response, using method-2 at 7 K and 300 K measured at low-Q is shown in Figs.14a-d. Fig. 14a reveals the absence of scattering below 5 meV and then the scattering rises and reaches a maximum near 13 meV, which is also supported from $E_i$=100 meV measurements (Fig. 14c). Further as observed in x=0.8, the 100 meV data also show the presence of a higher-energy peak (with $\Gamma\sim$ 22(1) meV) centred near 51 (1) meV at 7 K. It is to be noted that Q-dependence of the low energy peak 10-15meV in x=1 is very similar to that observed in x=0.8 and does not follow $F^2(Q)$ of Ce$^{3+}$ (due to weak magnetic intensity and strong phonon intensity in this case it was not possible to do any quantitative analysis of the Q-dependent intensity). On the other hand the energy integrated intensity of the 51 meV peak (in both x=1 and 0.8) exhibits $F^2(Q)$ behavior. Furthermore at 300 K, the spin gap response observed near 10-15 meV transforms into a quasi-elastic line (Figs.14b&d) and also the intensity of 51(2) meV peak decreases at 300 K. The estimated value of the susceptibility from these two INS peak is 1.4 x10$^{-3}$ (emu/mole), which is in good agreement with the measured dc-susceptibility (of the polycrystalline sample of x=1 ) 1.75 x 10$^{-3}$ (emu/mole) [13]. The low energy spin gap observed through INS study is also in agreement with 100 K (8.6 meV) gap estimated through the heat capacity measurements [13] and 125 K (11meV) gap estimated from the NMR measurements [21, 22], while the 51 meV energy scale is close to the 55 meV charge gap observed through the optical study [20]. To understand the origin of the 51 meV INS peak, we first compare the data with the observed INS response of x=0 (i.e CeRu$_2$Al$_{10}$). In x=0 in addition to the 8 meV spin gap, we have observed two well defined crystal electric field (CEF) excitations at 30 meV and 46 meV at 6 K. These excitations are temperature dependent and become broadened at 44 K [40]. This shows the observed broad excitation near 51(2) meV in x=0.8 and 1 could be interpreted as two broad CEF excitations in the presence of strong hybridization [47]. This type of change from two well defined CEF excitations to a broad hybridized response (note the difference from pure CEF excitations) has been observed in Ce(Ni$_{1-x}$Pt$_x$)Sn with the Pt composition x [47].

### (5) Spin gap as a function of Fe-composition (x)

At present there are no single crystal measurements available across the series of Ce(Ru$_{1-x}$Fe$_x$)$_2$Al$_{10}$, hence to compare the change in the spin gap energy with Fe composition (x) we have used the data from our powder samples. Figure 15 shows x dependence of the low energy spin gap estimated from INS data at 4.5 (and 7) K. It is clear that the gap is nearly constant up to x=0.5 and then increases towards high Fe-content compounds with x=0.8 and 1.0. It is interesting to note that x=0.8 and x=1 compounds are paramagnetic (PM) down to the lowest temperature. The presence of the larger gap in the PM compounds x=0.8 and 1 indicates that the gap is also related to the hybridization and not only due to the spin wave in Ce(Ru$_{1-x}$Fe$_x$)$_2$Al$_{10}$ (x=0 to 0.5). It is to be noted that the observed value of the gap (low energy gap) in x=1 is smaller than that expected using the scaling law discussed above and also in ref. [37]. At present we do not have any clear explanation, but this could be associated with the anisotropic nature of the gap in x=1.

### (6) Ce L$_3$-edge investigations

We have investigated the Ce L$_3$-edge x-ray absorption near-edge structure (XANES) of the x=1 material in order to shed further light on the origin of the broad scattering near 50 meV, that is, the CEF vs hybridization gap type of response in the INS data. . We have compared this data with that of x=0, where the Ce ion valence is nearly 3+, and the compound CeCoSi$_3$ in which the Ce ions are well-known to be in the mixed valence state [48]. The near edge structure of an x-ray absorption spectrum is sensitive to electronic transitions from the core level to the higher unfilled or half filled orbitals of the absorbing atom. XANES is therefore uniquely placed to measure valence states.

Fig. 16a shows the absorption spectra at 300 K from all three compounds and also the first-order energy derivative (Fig.16b) of these data. The figure shows that all three samples have a strong absorption peak at approximately 5728 eV, which corresponds to the 4f$^1$ state found in the Ce$^{3+}$ ions. Starting at approximately 5734 eV, the x = 1 and CeCoSi$_3$ data show a change in the slope that develops into a shoulder centred at around 5738 eV in the absorption spectra that is associated with the presence of $4f^0$ final states in the material and indicate the presence of Ce ions in the 4+ oxidation state [48]. The first order derivative data also confirm the different behavior of the three samples by showing the different rate of change in the absorption between the materials: faster for the x = 0 sample and slower for the x=1 and $CeCoSi_3$ samples. The observation of $Ce^{4+}$ indicates that the Ce ions are in a mixed valence state in the x = 1 material. As the probing time of the x-ray photons is much faster than the valence fluctuation time, when an electron jumps from $4f^1$ state to the conduction band (i.e. $Ce^{3+}$ ions with $4f^1$ becomes $Ce^{4+}$ ions with $4f^0$), the x-ray absorption study gives a snapshot of both valence states and we can observe the two features described above. This result indicates that the observed broad inelastic scattering at 50.8 meV in x=1 (and also x=0.8) is not due to pure CEF excitations, but to the hybridized 4f-conduction electron response as observed in $CePd_3$ [43], in other words the excitations across the lower and upper hybridization bands [4].

## IV. Discussion

The gap in the excitations spectrum (i.e absence of the quasi-elastic scattering) in x=1 was also confirmed through our low energy INS measurements on OSIRIS. This type of response has also been observed in our INS study of $CeCoSi_3$ (broad INS peak near 80 meV), while $CeTSi_3$ (T=Rh and Ir) exhibits well defined CEF excitations [49]. It is to be noted that single crystal susceptibility of $CeCoSi_3$ and also $CeRuSi_3$ (where Ce ions are also in a mixed valence state) exhibits anisotropy [50], which might have some relation to anisotropic hybridization and not pure CEF effect. This is also supported through the measured DC-susceptibility of $CeFe_2Al_{10}$ (single crystals), which is much smaller than that of $CeRu_2Al_{10}$ and almost half of that of $CeOs_2Al_{10}$ [35], supporting the role of anisotropic hybridization. This is also seen through the estimated moment values from the observed INS response using the moment sum rule of neutron scattering [41], which gives a paramagnetic moment, $\mu_{eff} = 1.4\,(3)\mu_B$ smaller than that expected of 2.54 $\mu_B$ for $Ce^{3+}$ ions. In the presence of strong hybridization, the INS response shifts towards high energy. So it is an open question whether the missing moment in x=1 is transferred to high energy or it is screened by the Kondo effect due to strong hybridization along the *b*-axis [20]. Further the analysis of the single crystal susceptibility of x=1 based on pure CEF model as presented in ref. [51] may not be the correct approach as the CEF ground state gives quasi-elastic (QE) scattering, but our low energy INS data has revealed the absence of QE scattering.

## V. Conclusions

We have carried out μSR and inelastic neutron scattering measurements on $Ce(Ru_{1-x}Fe_x)_2Al_{10}$ (x=0, 0.3, 0.5, 0.8 and 1) to understand the unusual magnetic phase transition and spin gap formation. Our μSR spectra of x=0, 0.3 and 0.5 clearly reveal the presence of two frequency oscillations below 27 K, which for the first time provides the direct evidence of the long-range magnetic ordering in these Fe partially substituted compounds. The temperature dependence of the μSR frequencies and the muon depolarization rates follow an unusual behavior with further cooling of the sample below 18 K, pointing to the possibility of another phase transition below 5 K. Further, the μSR spectra of x=0.8 and 1 do not provide any evidence of long-range magnetic ordering down to the lowest temperature confirming the paramagnetic ground state in these two compounds. The inelastic neutron scattering (INS) study has established the formation of a spin energy-gap with an energy scale of around 8 meV in the magnetic ordered compounds x=0, 0.3 and 0.5. Further INS results of x=0.5 show a possibility of another peak near 12 meV and we attribute two-peak type-structure to the dispersion of the excitations as seen in our data taken with $E_i$=100 meV. The temperature dependence of the inelastic peak position of x=0 and 0.3 reveals a possibility of existence of INS peak above $T_N$. More interestingly the INS results of paramagnetic compounds x=0.8 and 1 reveal the presence of inelastic peaks (or spin gap) at 10 and 12.5 meV, respectively and at high temperature the response transforms into a quasi-elastic line. The spin gap in these compounds are localised in Q-space (only seen at narrow Q-range), which may indicate that the origin of the spin gap is due to either gap opening on the small part of the Fermi-surface (anisotropic hybridization) or spin-dimer formation. Detailed μSR and neutron scattering measurements on a single crystal samples of x=0.8 and 1 are essential to understand the true nature of the spin gap, as we believe that magnetocrystalline anisotropy that is prevalent in $CeRu_2Al_{10}$ is likely to extend right across the substitutional series.

Our study has revealed important new results in the broader context of spin-gap formation driven by 4f- and conduction electrons hybridization. In particular, we have demonstrated how spin-gap formation can develop notwithstanding the existence of magnetic ordering (Fig. 15). This points to the operation of 4f electron spin with conduction electrons in two coexisting channels, but with very different outcomes: one is the development of long-range magnetic order that is mediated between spins by the conduction electron, while the other achieves hybridization-driven spin gap formation and works, in contrast, to the demise of the local moment. We believe that the coexistence of the Kondo semiconducting state with spin-gap formation and magnetic order to be unique among 4f-electron systems and it poses a perplexing new ground state for the strongly correlated class of materials.

**Acknowledgement:**

We would like to thank Drs Winfried Kockelmann and Aziz Daoud-Aladine for their help in neutron diffraction study on GEM and HRPD and Xpress Access beamtime on GEM was provided by the Science and Technology Facilities Council (STFC). We acknowledge interesting discussion with Drs Andrea Severing, Dimitry Khalyavin, Pascal Manuel, Vivek Anand, Amir Murani, Profs. Peter Riseborough, Qimiao Si and Piers Coleman. Some of us DTA/ADH would like to thank CMPC-STFC, grant number CMPC-09108, for financial support. The work at Hiroshima University was supported by a Grant-in-Aid for Scientific Research on Innovative

Area "Heavy Electrons" (20102004) of MEXT, Japan. AMS thanks the SA-NRF (Grant 78832) and UJ Research Committee for financial support. PP thanks the FRC of UJ for a doctoral study scholarship.

**Figure captions**

Fig.1 (color online) Neutron powder diffraction patterns of Ce(Ru$_{1-x}$Fe$_x$)$_2$Al$_{10}$ (x=0 to 1) from one of the detector banks of the GEM diffractometer at 300 K. The solid line through the experimental points represents the GSAS Rietveld refinement profile fit using space group Cmcm. The vertical short columns indicate the Bragg peak positions. The lowermost curve represents the difference between the experimental and calculated intensities.

Fig.2 (color online) (a) Magnetic ordering temperature versus Fe composition (x) of Ce(Ru$_{1-x}$Fe$_x$)$_2$Al$_{10}$ (x=0 to 1) alloys. The open circles are from ref. [35], solid down triangles from neutron diffraction ref. [25, and the present work], and red diamonds from the present μSR study. The open squares show the jump in the heat capacity at T$_N$, ΔC(T$_N$) from ref. [35]. (b) and (c) show the orthorhombic lattice parameters, *a*, *b*, *c* and (d) displays the unit cell volume (V) of Ce(Ru$_{1-x}$Fe$_x$)$_2$Al$_{10}$ (x=0 to 1) compounds as a function x (the solid lines are guide to the eye) .

Fig. 3 (color online) The interatomic distances versus Fe composition (x) of Ce(Ru$_{1-x}$Fe$_x$)$_2$Al$_{10}$ (x=0 to 1). For Ce-Al$_3$ there are two distances d1 (nearest) and d2 (second nearest)

Fig. 4 (color online) Zero-field μSR spectra plotted as asymmetry versus time at various temperatures for various Fe compositions (x) of Ce(Ru$_{1-x}$Fe$_x$)$_2$Al$_{10}$ (x=0 to 1). The solid lines depict fits using Eq.(2) for $T > 27K$ and Eq.(3) for $T < 27K$ (see text).

Fig. 5 (color online) Fit parameters of zero-field (ZF) μSR spectra of Ce(Ru$_{1-x}$Fe$_x$)$_2$Al$_{10}$ (x=0 to 1), muon precession frequencies vs temperature (left), and depolarization rate vs temperature (right). Two distinct frequencies have been found, leading also to two depolarization rates which are plotted in red and black symbols.

Fig. 6 (color online) (a & b) Neutron diffraction patterns of Ce(Ru$_{1-x}$Fe$_x$)$_2$Al$_{10}$ for x=0.5 at 5 K and 35 K (data scaled to 0.95 to match back ground) obtained from the OSIRIS spectrometer and (c-d) showing the 5 K data (symbols) with calculated magnetic intensity (line) with the Ce moment of 0.17(4)μ$_B$ along c-axis using Fullprof programme.

Fig.7 (color online) (a) The integrated intensity of (101) diffraction peak and (b) background as a function of temperature of Ce(Ru$_{1-x}$Fe$_x$)$_2$Al$_{10}$ x=0.5.

Fig. 8 (color online) Colour coded inelastic neutron scattering intensity of Ce(Ru$_{1-x}$Fe$_x$)$_2$Al$_{10}$ (x=0, 0.3 and 0.5) and LaFe$_2$Al$_{10}$, at 5 K measured with respective incident energies of E$_i$=20 and 25 meV on the MARI spectrometer.

Fig. 9 (color online) Q-integrated intensity versus energy transfer of $Ce(Ru_{1-x}Fe_x)_2Al_{10}$ (x=0, 0.3 and 0.5) and $LaRu_2Al_{10}$, at 4.5 K measured with respective incident energies of $E_i$=20 (for x=0 and 0.3) and 25 meV (for x=0.5) on the MARI spectrometer.

Fig. 10 (color online) Q-integrated magnetic scattering intensity versus energy transfer of $Ce(Ru_{1-x}Fe_x)_2Al_{10}$ for x=0 (left), 0.3 (middle) and 0.5 (right), at different temperatures at Q=1.27 Å$^{-1}$. The solid line represents the fit using an inelastic peak (dash-dotted line represents the components of fit) and the dotted line represents the fit using a quasi-elastic peak (line above $T_N$, in the left bottom figure, black dotted line for 33 K and green dotted line for 40 K).

Fig.11 (color online) The fit parameters, susceptibility, linewidth and peak position versus temperature obtained from fitting the magnetic scattering intensity of $Ce(Ru_{1-x}Fe_x)_2Al_{10}$ (x=0 and 0.3). For comparison purposes we have also plotted the data of x=0 from Ref. [28] using open squares. The closed circles represent the fit using an inelastic peak and open circles represent the fit using a quasi-elastic peak. The small blue circles in (a) shows the measured dc-susceptibility of the polycrystalline sample of x=0 from ref. [13] and in (b) the dotted and solid line represents the fits using exponential and $T^2$ behavior respectively (see text). In (c) the dotted line represents the theoretical predicted behavior of the spin-spiral gap by Hanzawa [31] and the open triangles are for $N_{eff}^{1/2} \sim \Delta_{op}$ from ref. [19].

Fig. 12 (color online) Color coded inelastic neutron scattering intensity of $Ce(Ru_{1-x}Fe_x)_2Al_{10}$ x=0.8 (a) at 7 K and (b) at 94 K and of $LaFe_2Al_{10}$ at (c) at 7 K and (d) at 94 K measured with an incident energy of $E_i$=40 on the MERLIN spectrometer. The estimated magnetic scattering, after subtracting the phonon contribution, is shown in (e) at 7 K and (f) at 94 K.

Fig. 13 (color online) The estimated magnetic scattering of $Ce(Ru_{1-x}Fe_x)_2Al_{10}$ with x=0.8 (a) at 7 K and (b) at 94 K by taking the Q-integrated (1.01 Å$^{-1}$) energy cuts form Fig. 12 (shown by open circles). To further confirm the presence of magnetic scattering, we have used second method (method-2, see text) to estimate the magnetic scattering (blue squares): (c)  The Q-dependent energy integrated (8-12 meV) intensity of x=0.8 and $LaFe_2Al_{10}$  at 7 K and (d) the Q-dependent magnetic scattering (8-12meV) of x=0.8 at 7 K. The solid line represents the $Ce^{3+}$ magnetic form factor squared from Ref. [52] and the red dotted line represents the fit based on an isolated dimer structure factor (see text).

Fig. 14 (color online) The estimated magnetic scattering of x=1 (a) at 7 K and (b) at 300 K using an incident energy of 20 meV and (c) at 7 K and (d) at 300 K using an incident energy of 100 meV (also the data of 20 meV are plotted by open circles). The magnetic scattering was estimated using the method-2 (see text).  The solid line shows the fit to a Lorentzian function and dotted line shows the components of the fit. It is to be noted that in (c and d) 100 meV data are scaled to match 20 meV data due to form factor and background differences.

Fig. 15  (color online) The Fe composition (x) dependent of the inelastic peak position (lower energy peak, assigned to the spin gap, right y-axis) of $Ce(Ru_{1-x}Fe_x)_2Al_{10}$  at 4.5 K (and 7 K) estimated using MARI or MERLIN spectrometer. The left y-axis shows the $T_N$ (estimated from µSR) versus x.

Fig. 16 Ce (a) $L_3$-edge XANES spectra of $CeRu_2Al_{10}$ (blue solid line), $CeFe_2Al_{10}$ (red dash-dotted line) and the reference $CeCoSi_3$ (black dotted line) at room temperature and (b) the first order energy derivative.

**Appendix:**

Fig. 1A (top) The estimated magnetic scattering of $Ce(Ru_{1-x}Fe_x)Al_{10}$ with x=0 plotted as energy transfer (E) versus momentum transfer (Q) at 4.5 K measured on MARI using $E_i$=20 meV. The arrow indicates the position of (0 1 0) magnetic reflection at Q=0.61 Å$^{-1}$. (bottom) The Q-dependent energy integrated (7-10 meV) magnetic intensity of $CeRu_2Al_{10}$ at 4.5 K. The solid line represents the $Ce^{3+}$ magnetic form factor squared from Ref. [52] and the red dotted line represents the fit based on an isolated dimer structure factor (see text).

Fig. 2A The estimated magnetic scattering of $Ce(Ru_{1-x}Fe_x)Al_{10}$ with x=0.5 plotted as energy transfer (E) versus momentum transfer (Q) at 4.5 K measured on MARI using $E_i$=100 meV. The phonon scattering was subtracted by taking the average of $LaRu_2Al_{10}$ and $LaFe_2Al_{10}$ data. Dispersive excitations can be seen at 8-14 meV. (bottom) Intensity versus energy transfer at two different Q-positions indicating the presence of the dispersions.

**References:**

[1] C.M. Varma, Rev. Mod. Phys. **48**, 219 (1976)

[2] M. B. Maple, R.E. Baumbach, N. P. Butch, J.J. Hamlin and M.Janoschek, J. Low Temp. Phys. **161**, 4 (2010)

[3] A. Georges, G. Kotliar, W. Krauth and M.J. Rozenberg, Rev. Mod. Phys. **68**, 13 (1996)

[4] P.S. Riseborough, Adv. Phys. **49**, 257 (2000); P.S. Riseborough, Phys. Rev. B **45**, 13984 (1992)

[5] P. Coleman, in "Handbook of Magnetism and Advanced Magnetic Materials," ed. by H. Knoemuller and S. Parkin, John Wiley and Sons, Vol.**1**, 95 (2007)

[6] G.R. Stewart, Rev. Mod. Phys. **73**, 797 (2001)

[7] A. Amato, Rev. Mod. Phys. **69,** 1119 (1997)

[8] H. v. Löhneysen, A. Rosch, M. Vojta, and P. Woelfle, Rev. Mod. Phys. **79**, 1015 (2007)

[9] K. Takegahara, H. Harima, Y. Kaneta, A. Yanase, J. Phys. Soc. Jpn. **62**, 2103 (1993)

[10] L.F. Matheiss, D.R. Hamman, Phys. Rev. B **47**, 13114 (1993)

[11] A.M. Strydom, Physica B **404**, 2981 (2009)

[12] Y. Muro. J. Kajino, K. Umeo, K. Nishimoto and R. Tamura and T. Takabatake, Phys. Rev. B **81**, 214401(2010**);** K. Yutani, Y. Muro, J. Kajino, T. J. Sato and T. Takabatake, J.Phys. Conf. Ser., **391**, 012070 (2012)

[13] Y. Muro, K. Motoya, Y. Saiga and T. Takabatake, J. Phys.: Conf. Ser. **200**, 012136 (2010); Y. Muro, K. Motoya, Y. Saiga, and T. Takabatake, J. Phys. Soc. Jpn. **78**, 083707 (2009)

[14] T. Takasaka, K. Oe, R. Kobayashi, Y. Kawamura, T. Nishioka, H. Kato, M. Matsumura, and K. Kodama, J. Phys.: Conf. Ser. **200**, 012201 (2010); M. Matsumura, Y. Kawamura, S. Edamoto, T. Takesaka, H. Kato, T. Nishioka, Y. Tokunaga, S. Kambe, and H. Yasuoka, J. Phys. Soc. Jpn. **78** 123713 (2009)

[15] V.M. T. Thiede, T. Ebel, and W. Jeitschko, J. Mater. Chemistry, **8**, 125 (1998); A.I. Tursina, S. N. Nesterenko, E. V. Murashova, H. V. Chernyshev, H. Nöel, and Y. D. Seropegin,, Acta Crystallogr. Sect. E**61**, 112 (2005)

[16] T. Nishioka, Y. Kawamura, T. Takesaka, R. Kobayashi, H. Kato, M. Matsumura, K. Kodama, K. Matsubayashi, and Y. Uwatoko: J. Phys. Soc. Jpn. **78** 123705 (2009)

[17] H. Tanida, D. Tanaka, M. Sera, C. Moriyoshi, Y. Kuroiwa, T. Nishioka, H. Kato, and M. Matsumura, J. Phys. Soc. Jpn. **79**, 043708 (2010)

[18] S. Kimura, T. Iizuka, H. Miyazaki A. Irizawa, Y. Muro and T. Takabatake, Phys. Rev. Lett. **106,** 056404 (2011)

[19] S. Kimura, T. Iizuka, H. Miyazaki, T. Hajiri, M. Matsunami, T. Mori, A. Irizawa, Y. Muro J. Kajino, and T. Takabatake, Phys. Rev. B **84**, 165125 (2011)

[20] S. Kimura, Y. Muro, and T. Takabatake, J. Phys. Soc. Jpn. **80**, 033702 (2011).

[21] S.C. Chen and Lue, Phys. Rev. B **81**, 075113 (2010)

[22] Y. Kawamura, S. Edamoto, T. Takesaka, T. Nishioka, H. Kato, M. Matsumura, Y. Tokunaga, S. Kambe, and H. Yasuoka: J. Phys. Soc. Jpn. **79**, 103701 (2010)

[23] T. Takabatake, F. Teshima, H. Fujii, S. Nisjhigori, T. Suzuki, T. Fujita, Y. Yamaguchi, J. Sakurai, and D. Jaccard, Phys. Rev. B **41**, 9607 (1990)

[24] S.K. Malik and D.T. Adroja, Phys. Rev. B **43**, 6277 (1991)

[25] D. Khalyavin, A. D. Hillier, D. T. Adroja, A. M. Strydom, P.Manuel, L. C. Chapon, P. Peratheepan, K. Knight, P. Deen, C. Ritter, Y. Muro, and T. Takabatake: Phys. Rev. B **82**, 100405(R) (2010)

[26] S. Kambe, H. Chudo, Y. Tokunaga, T. Koyama, H. Sakai, U. Ito, K. Ninomiya, W. Higemoto, T. Takesaka, T. Nishioka and Y. Miyake, J. Phys. Soc. Jpn, **79**, 053708 (2010)

[27] D. T. Adroja, A. D. Hillier, P. P. Deen, A. M. Strydom, Y. Muro, J. Kajino, W. A. Kockelmann, T. Takabatake, V. K. Anand, J. R. Stewart, and J. Taylor, Phys. Rev. B **82** 104055 (2010)

[28] J. Robert, J.M. Mignot, G. André, T. Nishioka, R. Kobayashi, M. Matsumura, H. Tanida, D. Tanaka and M. Sera, Phys. Rev. B **82** 100404 R (2010); J. M. Mignot et al J. Phys. Soc. Jpn (2011)

[29] A.M. Strydom, D.T. Adroja, P. Deen, C. Ritter, ILL Experimental Report (2010); D. Khalyavin, D. T. Adroja, A. M. Strydom and P. Manuel, unpublished (2013)

[30] H. Kato, R. Kobayashi, T. Takesaka, T. Nishioka, M. Matsumura, K. Kaneko, and N. Metoki, J. Phys. Soc. Jpn, Supplement **80**, 073701 (2011).

[31] K. Hanzawa, J. Phys. Soc. Jpn. **79**, 043710 (2010)

[32] K. Hanzawa, J. Phys. Soc. Jpn **80**, 113701 (2011)

[33] D.T. Adroja, T. Takabatake and Y.Muro, ISIS Facility experimental report, RB1110535 (2012)

[34] J. Robert, J-M Mignot, S. Petit, P. Steffens,T. Nishioka, R. Kobayashi,M. Matsumura, H. Tanida, D. Tanaka, and M. Sera, Phys. Rev. Lett. 109, 267208 (2012)

[35] T. Nishioka, D. Hirai, Y. Kuwamura, H. Kato, M. Matsumurata, H. Tanida, M. Sera, K. Matsubayashi and Y. Uwatoko, .J. Phys.: Conf. Ser. **273,** 012046 (2011)

[36] D.T. Adroja, J.-G. Park, E.A. Goremychkin, K.A. McEwen, N. Takeda, B.D. Rainford, K.S. Knight, J.W. Taylor, J. Park, H.C. Walker, R. Osborn, and P. S. Riseborough , Phys. Rev. B **75**, 014418 (2007)

[37] D. T. Adroja, K.A. McEwen. J.-G. Park, A.D. Hillier, N. Takeda, P.S. Riseborough and T. Takabatake, J. Opto. Adv. Mate., **10**, 1564 (2008)

[38] F. Strigari, T. Willers, Y. Muro, K. Yutani, T. Takabatake, Z. Hu, Y.-Y. Chin, S. Agrestini, H.-J. Lin, C. Chen, A. Tanaka, M. W. Haverkort, L.-H. Tjeng, and A. Severing, Phys. Rev. B **81**, 081105(R) (2012).

[39] R. Kubo, T. Toyabe, in "Magnetic Resonanceand Relaxation," ed. by R. Blinc, North-Holland, Amsterdan, p. 810 (1966); R.S. Hayano, Y.J. Uemura, J. Imazato, N. Nishida, T. Yamazaki and R. Kubo, Phys. Rev. B **20**, 850 (1979)

[40] D.T. Adroja, P.P. Deen, A.D. Hillier, Y. Muro, T. Takabatake and A. M. Strydom, unpublished (2013)

[41] D. T. Adroja, J.-G. Park, K. A. McEwen, N. Takeda, M. Ishikawa, J.-Y. So, Phys. Rev. B **68**, 094425 (2003);

[42] H. Güdel and A. Furrer, Mol. Phys. **33**, 1335 (1977)

[43] A.P. Murani, A. Severing, W.G. Marshall, Phys. Rev. B **53**, 2641 (1996)

[44] R.Viennois, L. Girard, L.C. Chapon, D.T. Adroja, R.I. Bewley, D. Ravot, P. S. Riseborough, and S. Paschen, Phys. Rev. B **76**, 174438 (2007)

[45] A.J. Fedro and S.K. Sinha, in Valence Fluctuations in Solids, edited by L.M. Falicov, W. Hanke, and M.B. Maple (North-Holland, Amsterdam, p. 329 (1981)

[46] T.E. Mason, G. Aeppli, A.P. Ramirez, K.N. Clausen, C. Broholm, N. St¨ucheli, E. Bucher and T.I.M. Palstra, Phys. Rev. Lett. 69, 490 (1992); T.J. Sato, H. Kadowaki, H. Yoshizawa, T. Ekino, T. Takabatake, H. Fuji, L.P. Regnault and Y. Isikawa, J. Phys. Condens. Matter 7, 8009 (1995)

[47] D. T. Adroja, B. D. Rainford, A. J. Neville, and A. G. M. Jansen, Physica B **223&224**, 275 (1996)

[48] P. Le Fèvre, H. Magnan, J. Vogel, V. Formoso, K. Hricovini and D. Chandesris, J. Synchrotron Rad. **6**, 290 (1999)

[49] D.T. Adroja, A.D. Hillier, V.K. Anand et al unpublished (2013)

[50] T. Kawai, H. Murankai, M. A. Meassoni, T. Shimodai, Y. Doii,T. D. Matsuda, Y. Haga, G. Knebel, G. Lapertot, D. Aoki, J. Flouquet, T. Takeuchi, R. Settai, and Y. Ŏnuki, J. Phys. Soc. Jpn. **77**, 064716 (2008).

[51] F. Strigari, T. Willers, Y. Muro, K. Yutani, T. Takabatake,Z. Hu, S. Agrestini, C.-Y. Kuo, Y.-Y. Chin, H.-J. Lin, T. W. Pi, C. T. Chen, E. Weschke, E. Schierle, A. Tanaka, M. W. Haverkort, L. H. Tjeng, and A. Severing, Phys. Rev. B87, 125119 (2013)

[52] P.J. Brown, in International Tables for Crystallography, edited by A.J.C. Wilson, Mathematical, Physical and Chemical Tables Vol. C (Kluwer Academic, Amsterdam, 1999), pp. 450–457.

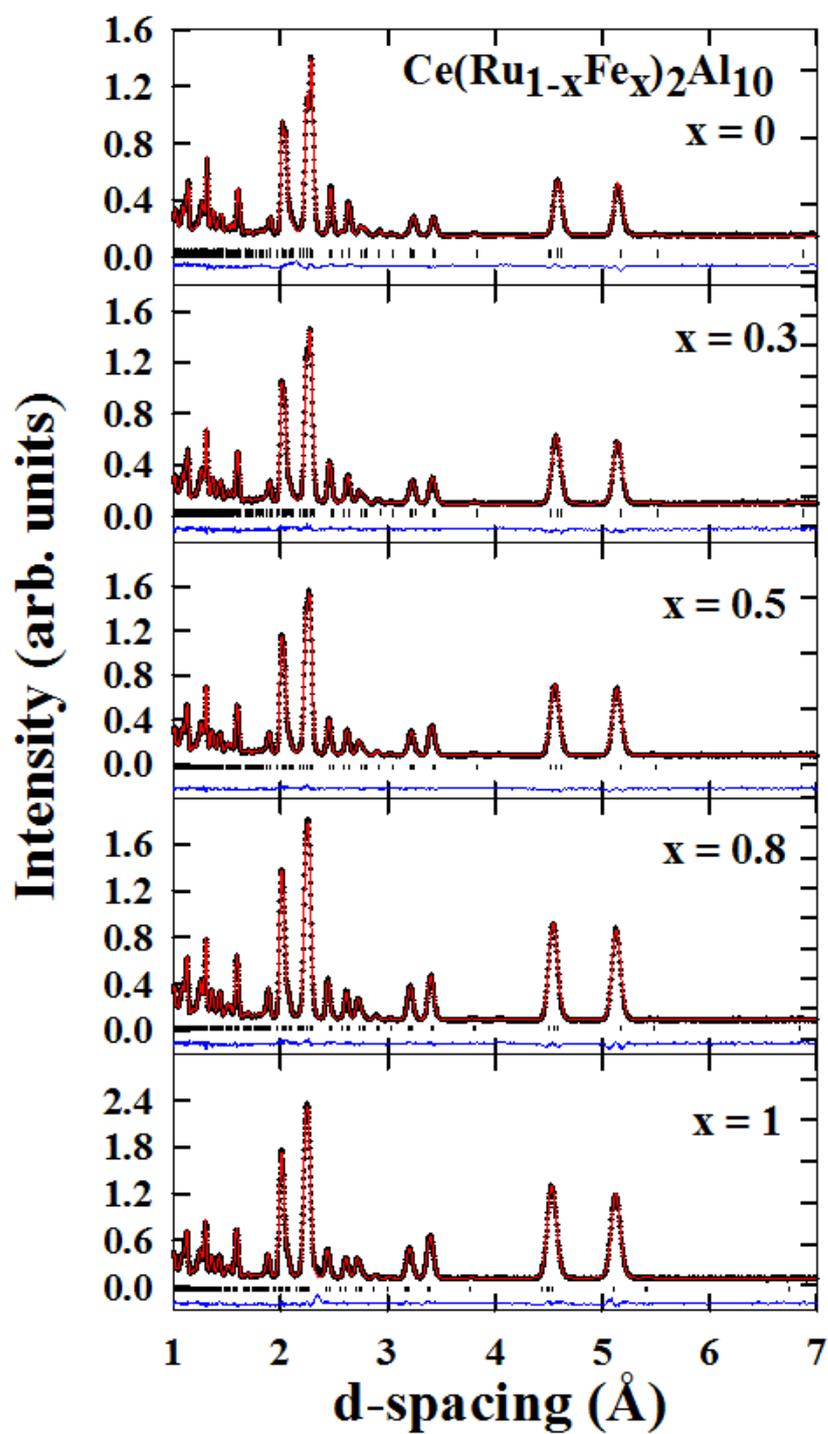

Fig.1 Adroja et al

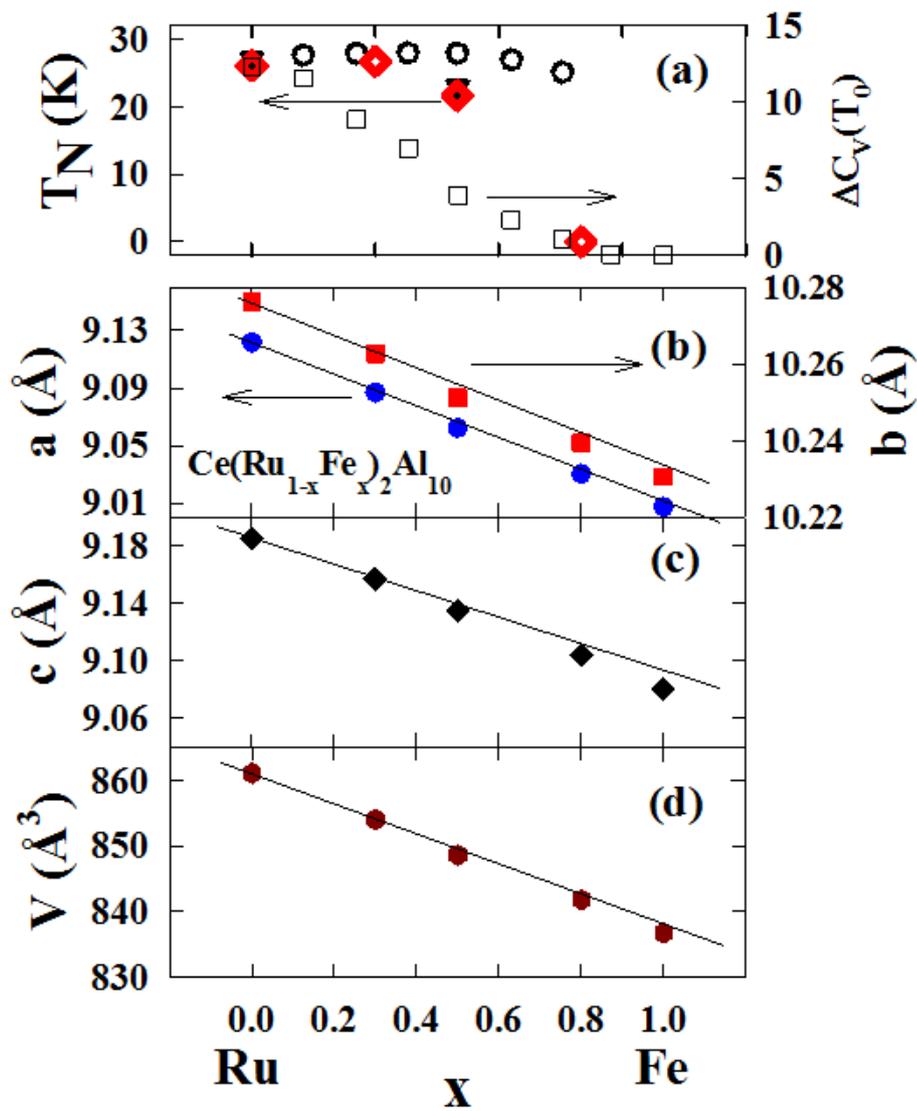

Fig. 2 Adroja et al

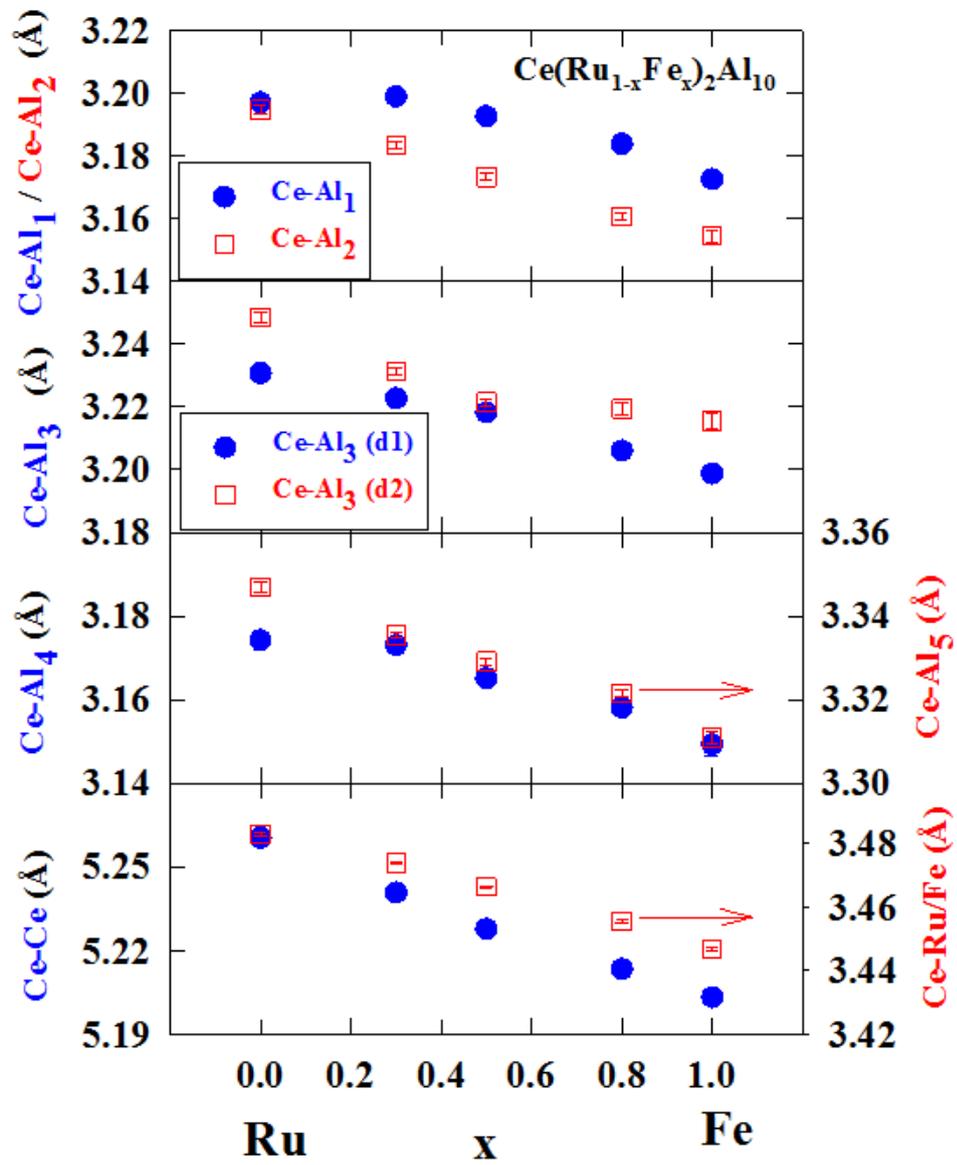

Fig.3 Adroja et al

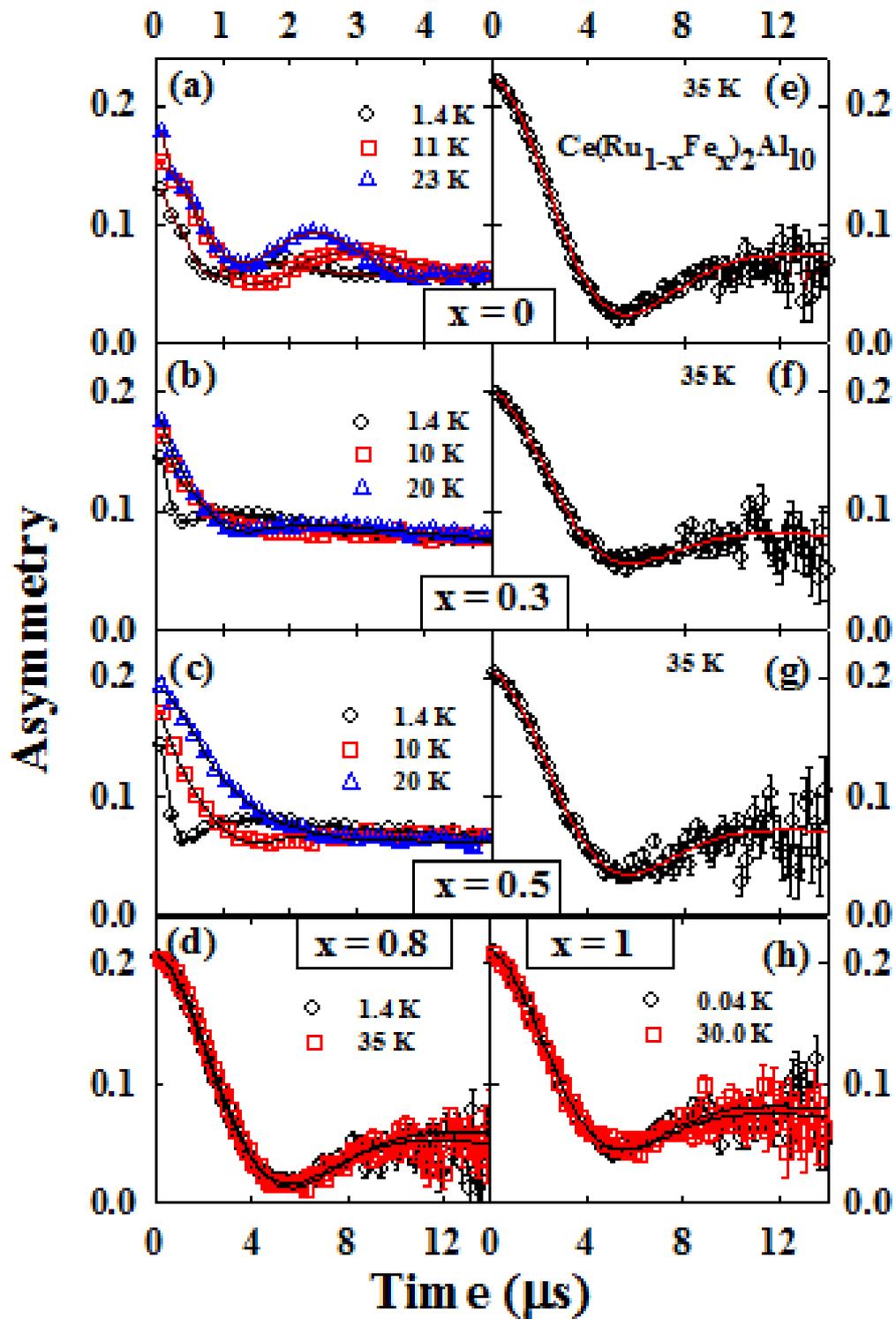

Fig.4 Adroja et al

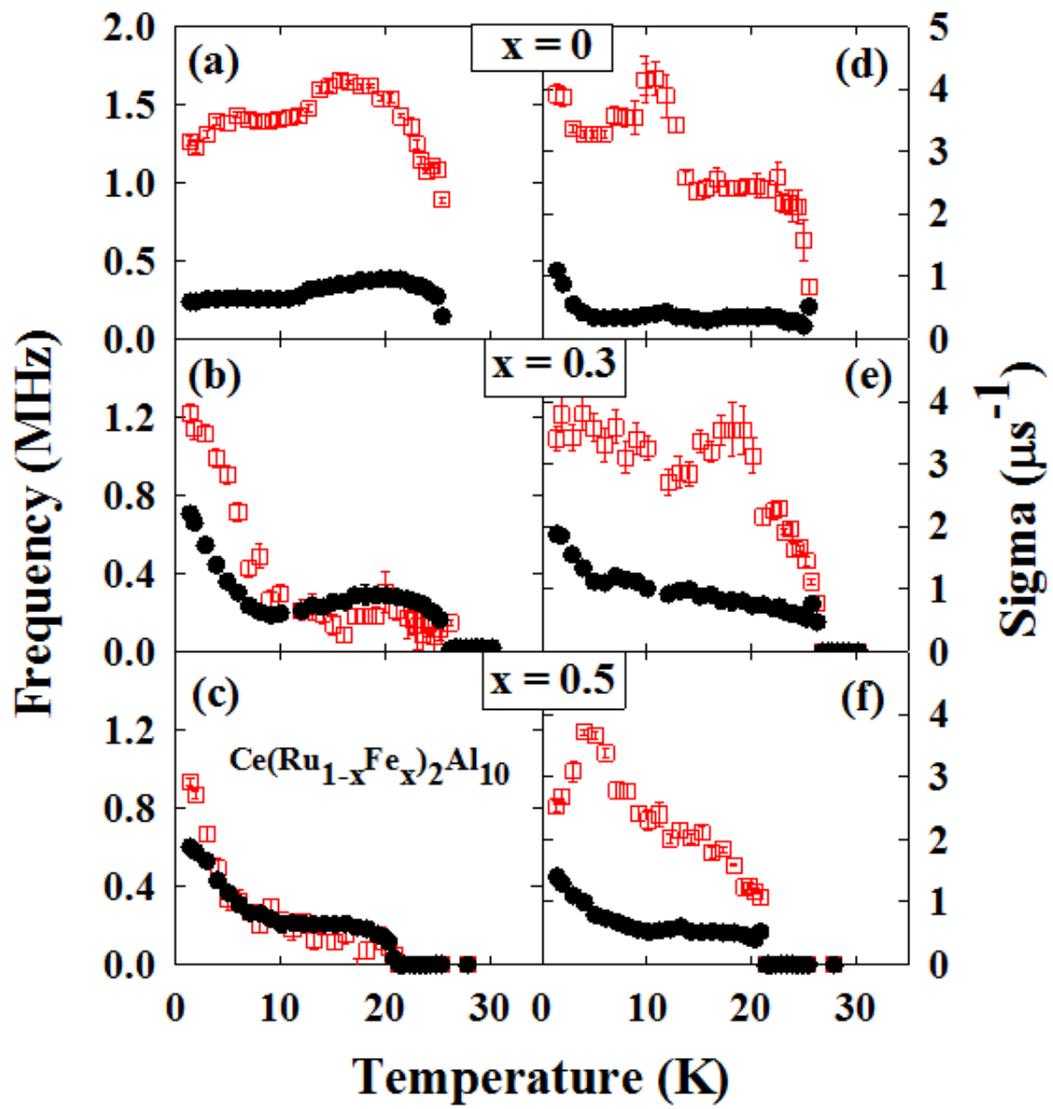

Fig.5 Adroja et al

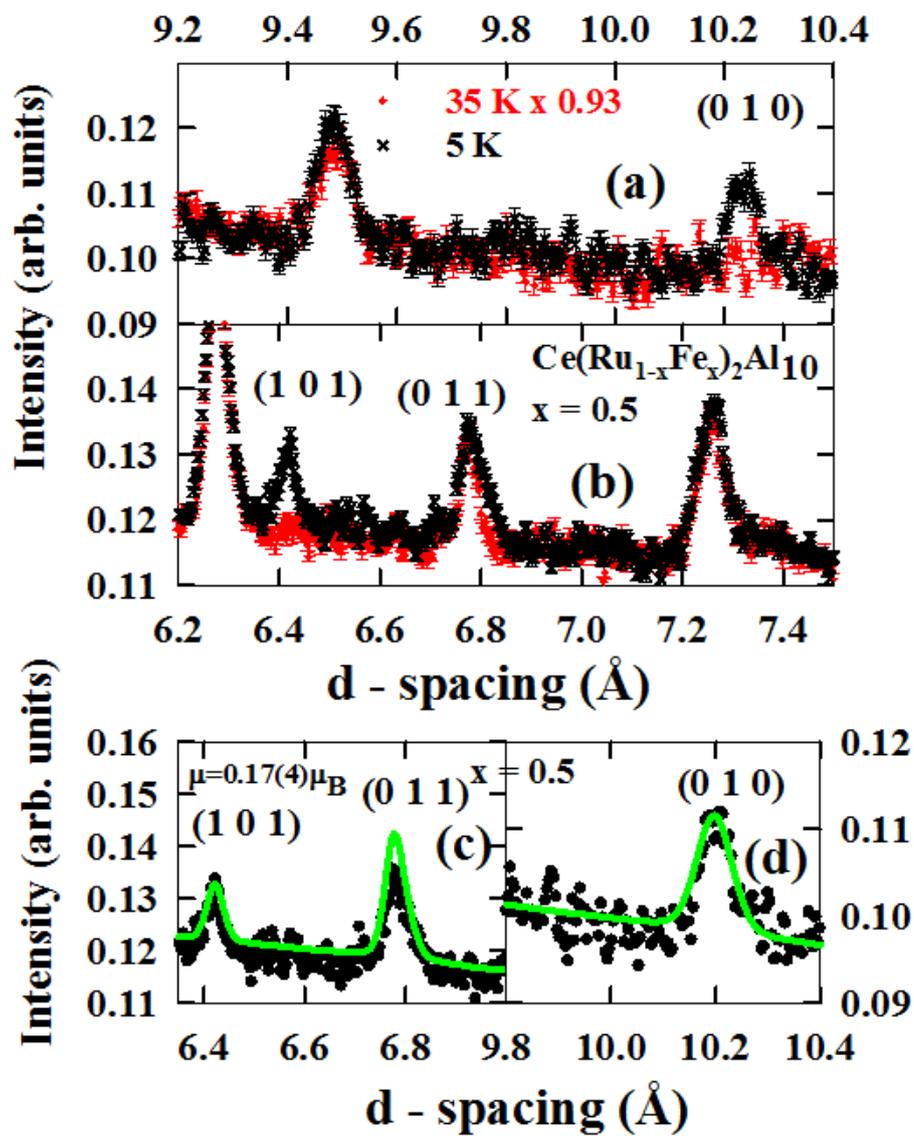

Fig. 6 Adroja et al

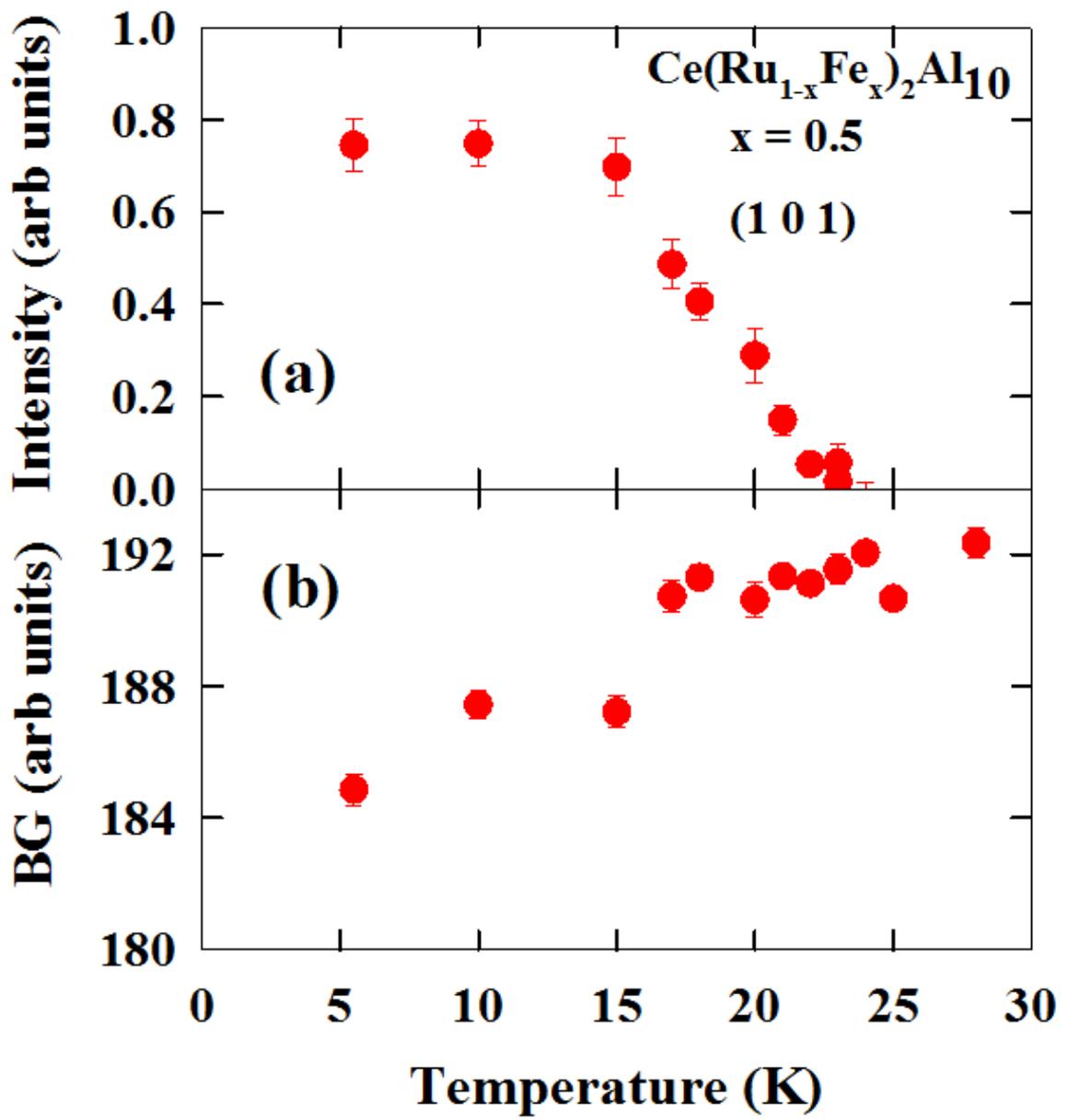

Fig.7 Adroja et al

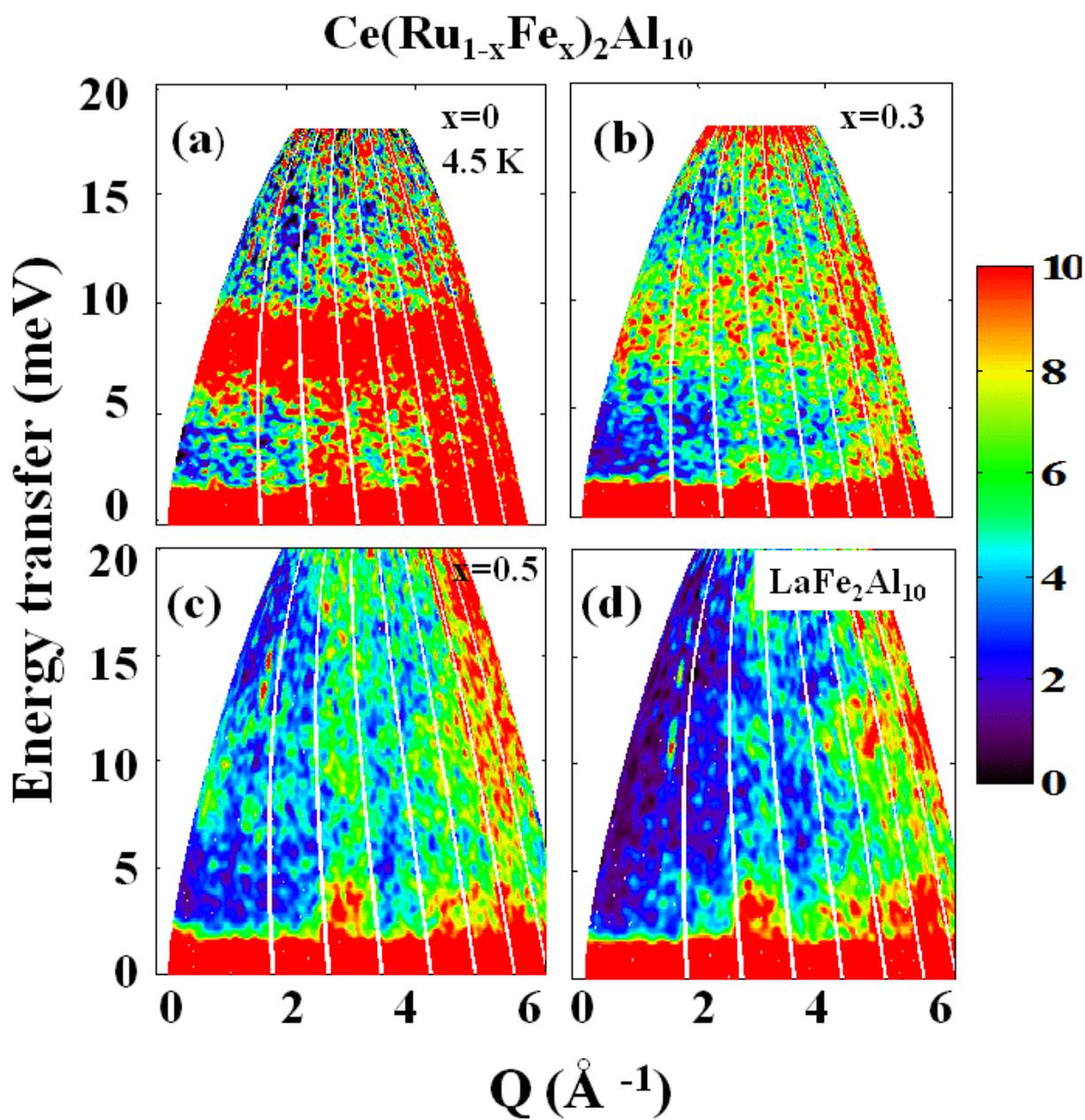

Fig. 8 Adroja et al

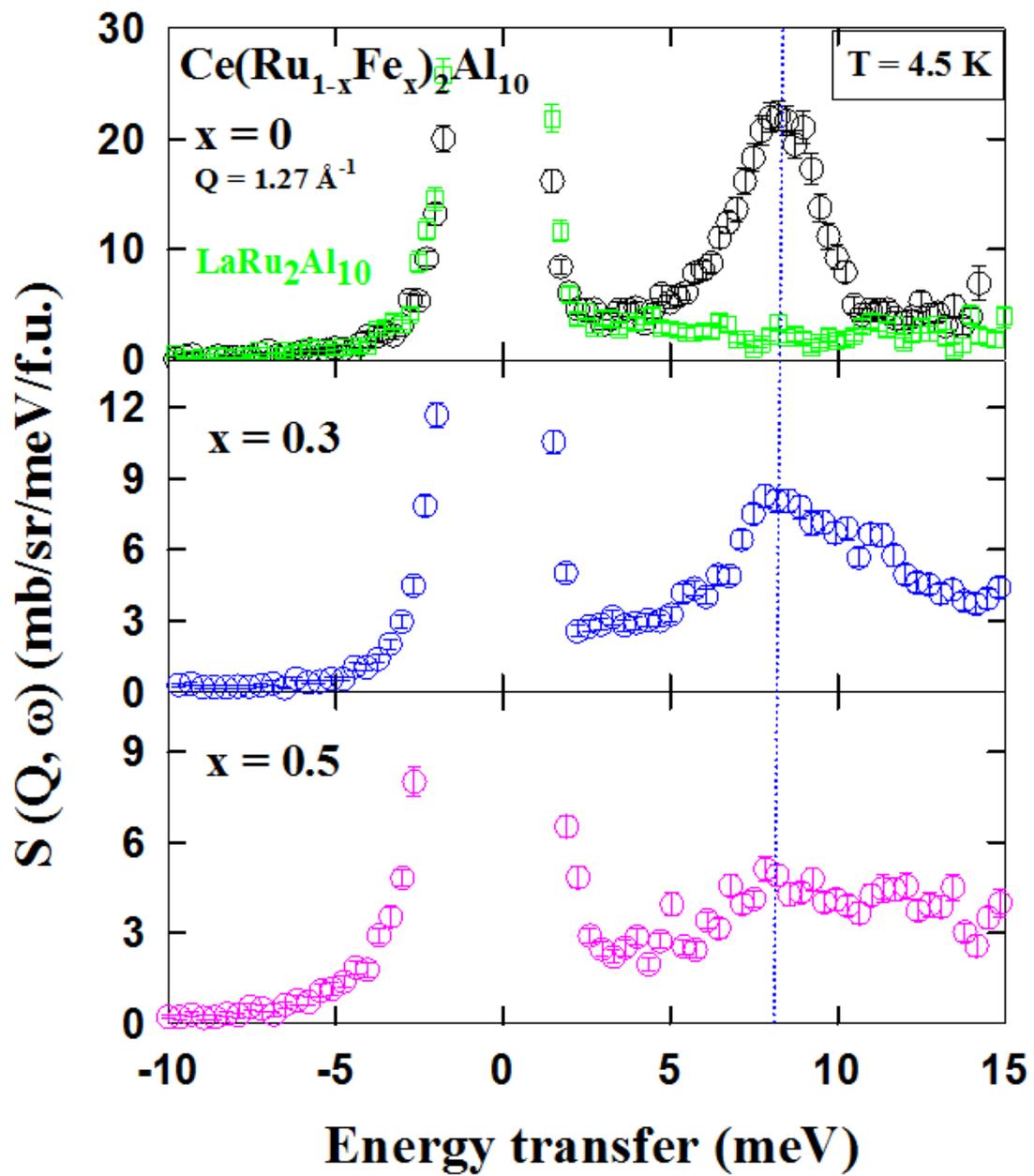

Fig.9 Adroja et al

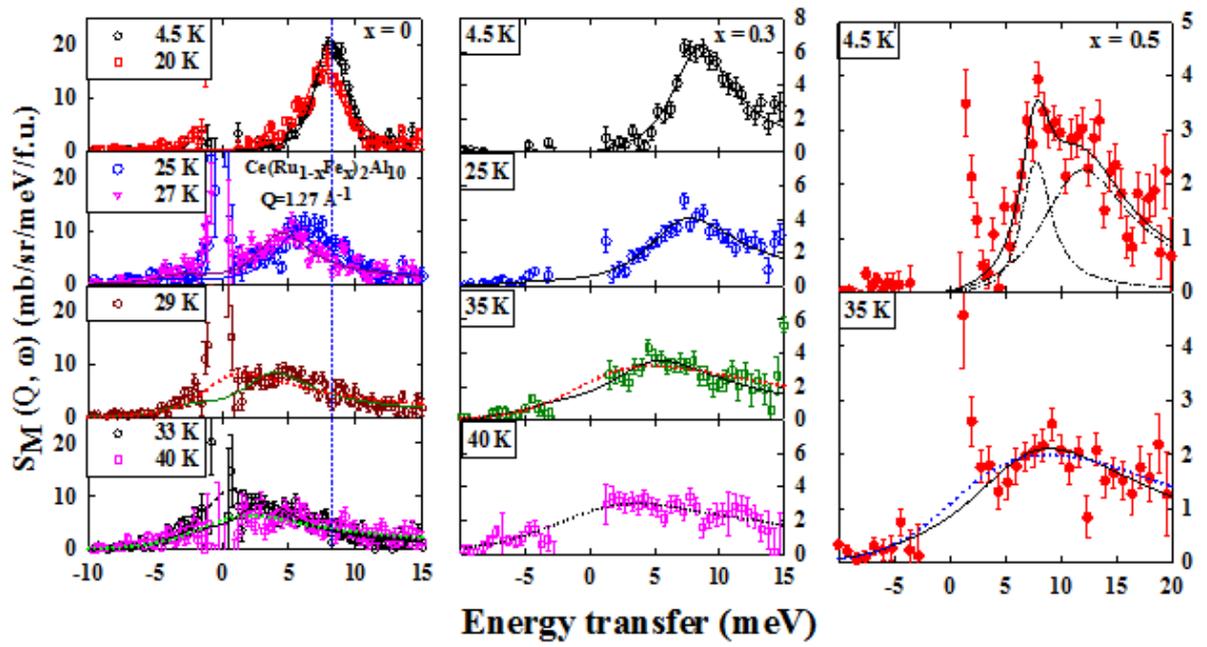

Fig. 10 Adroja et al

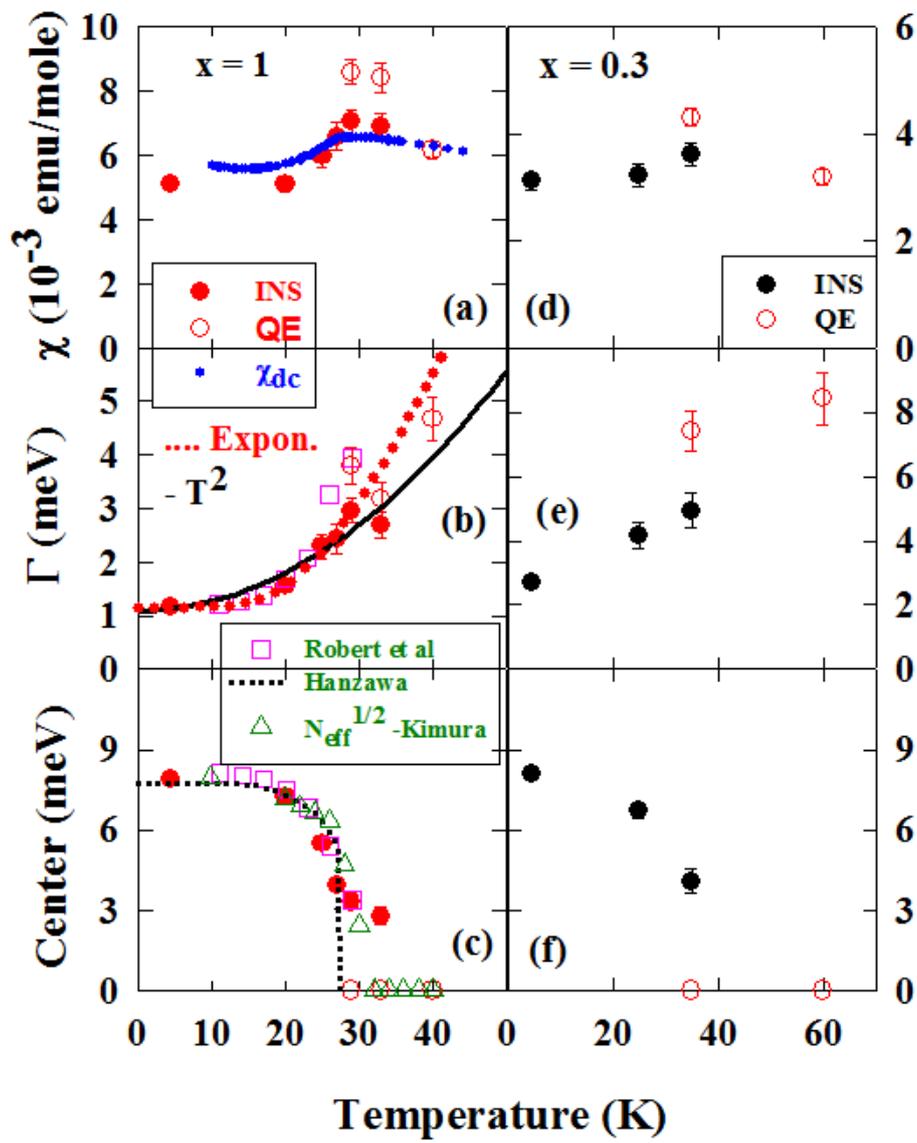

Fig. 11 Adroja et al

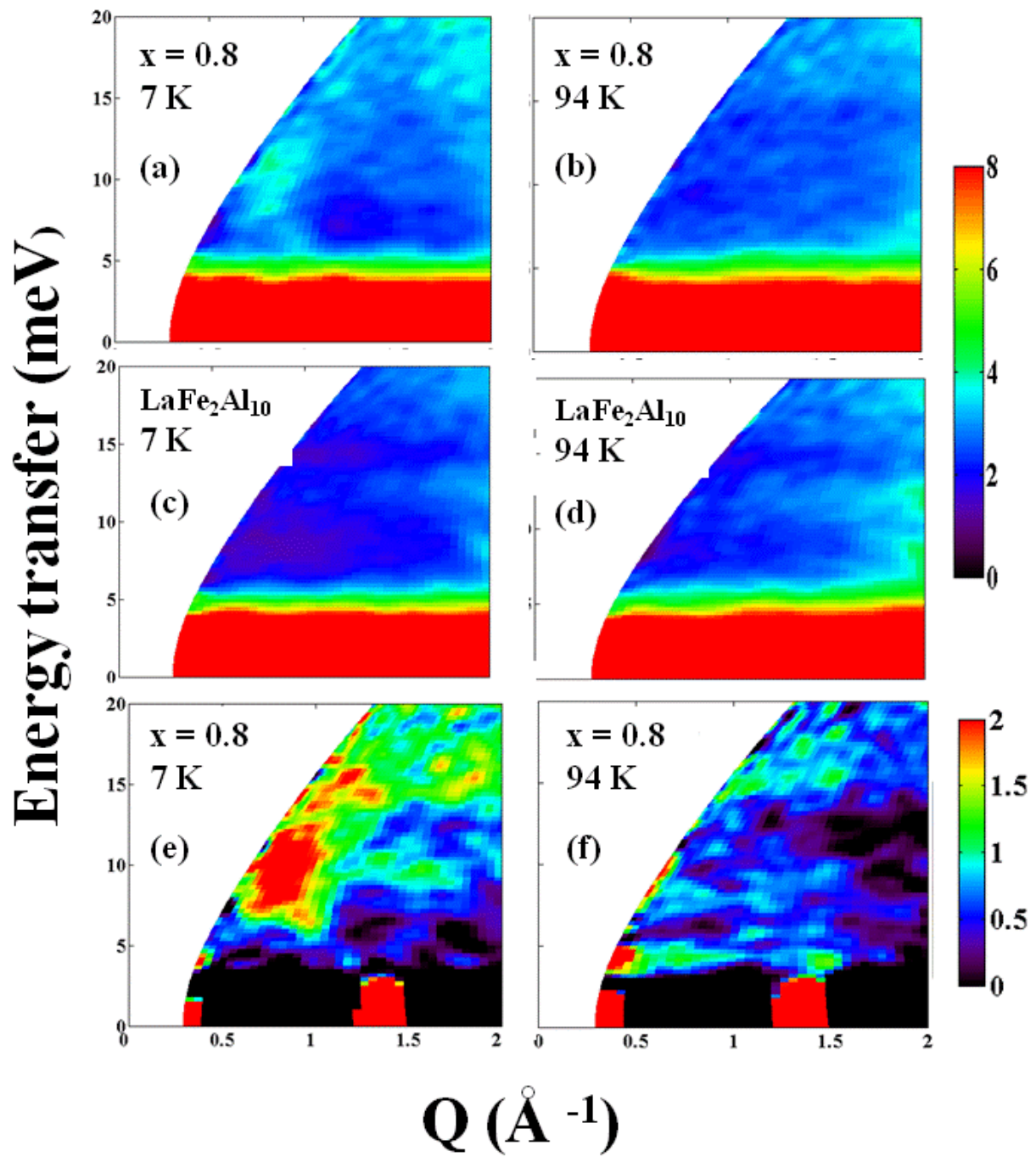

Fig. 12   Adroja et al

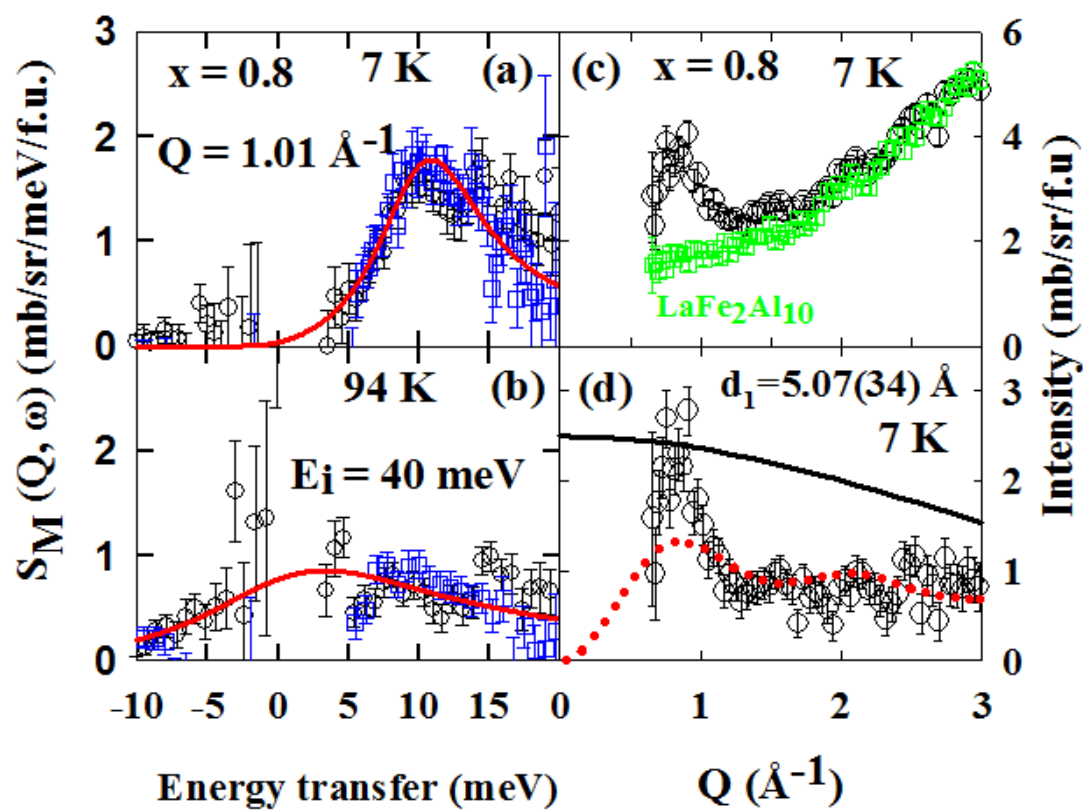

Fig. 13 Adroja et al

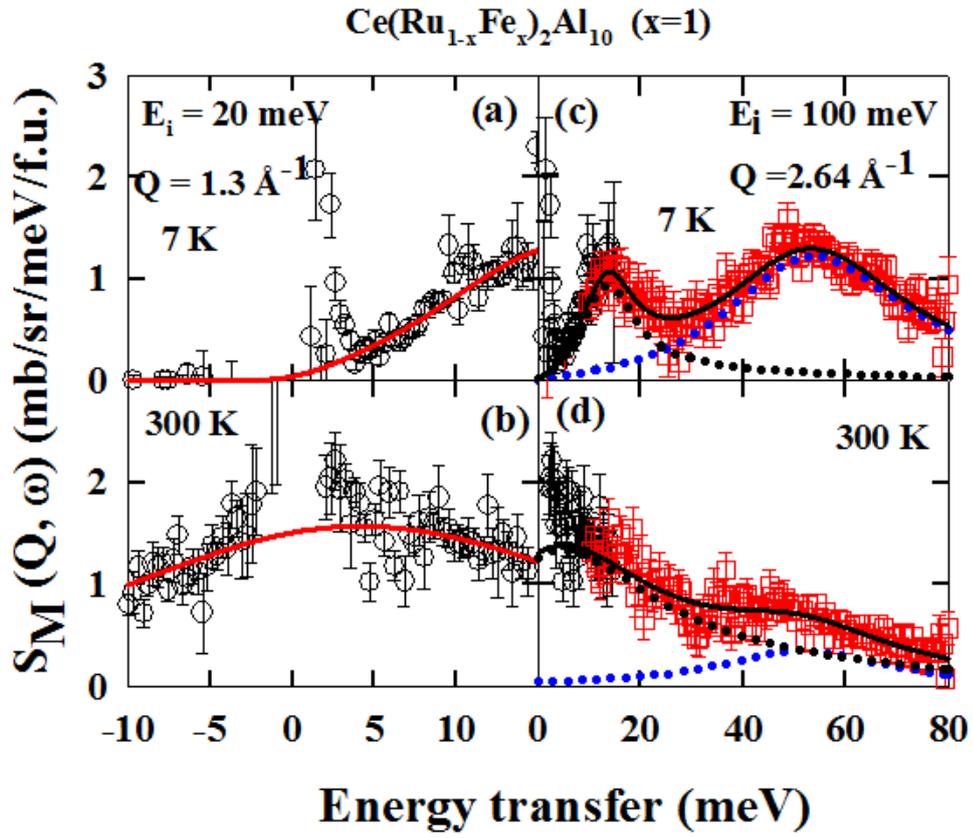

Fig. 14 Adroja et al

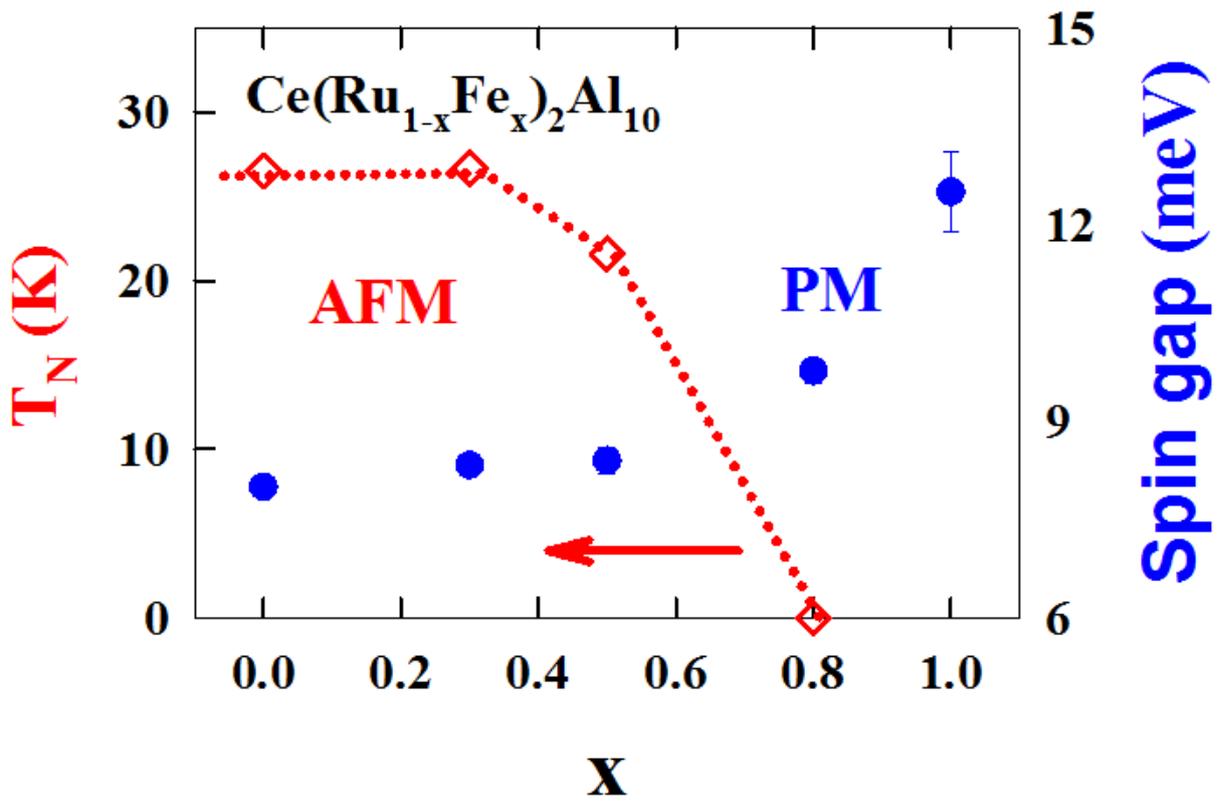

Fig.15 Adroja et al

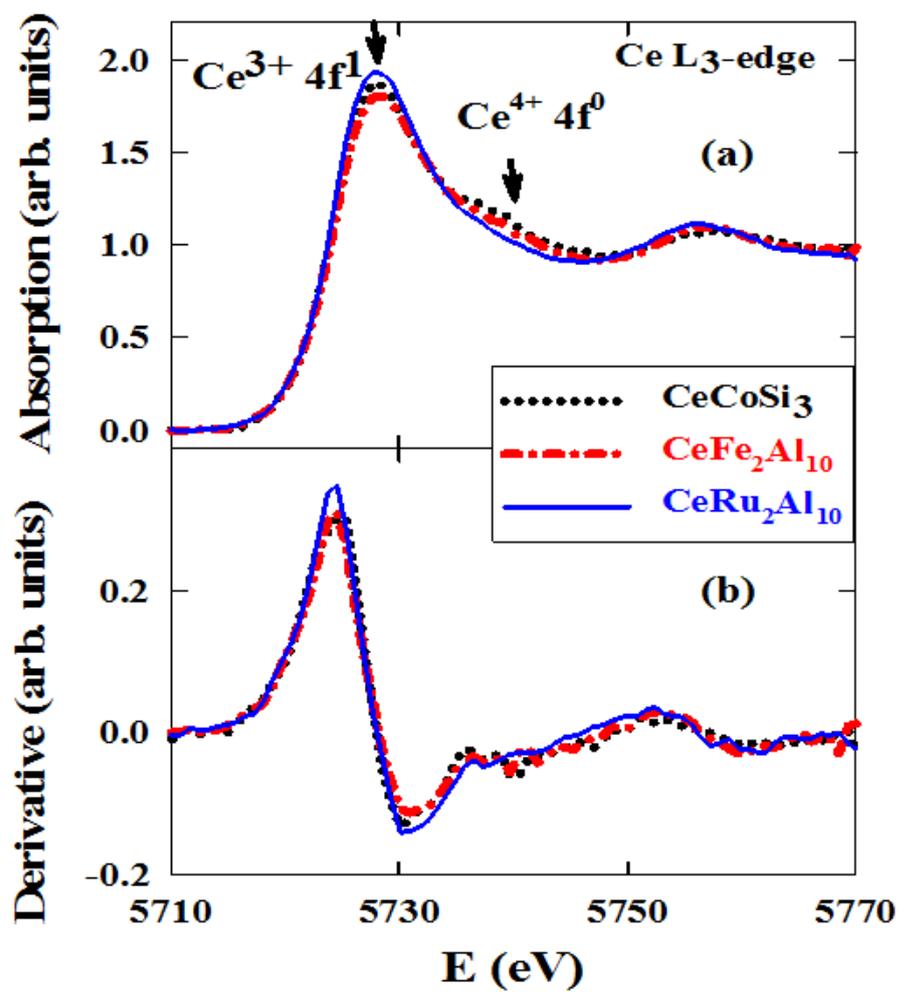

Fig.16 Adroja et al

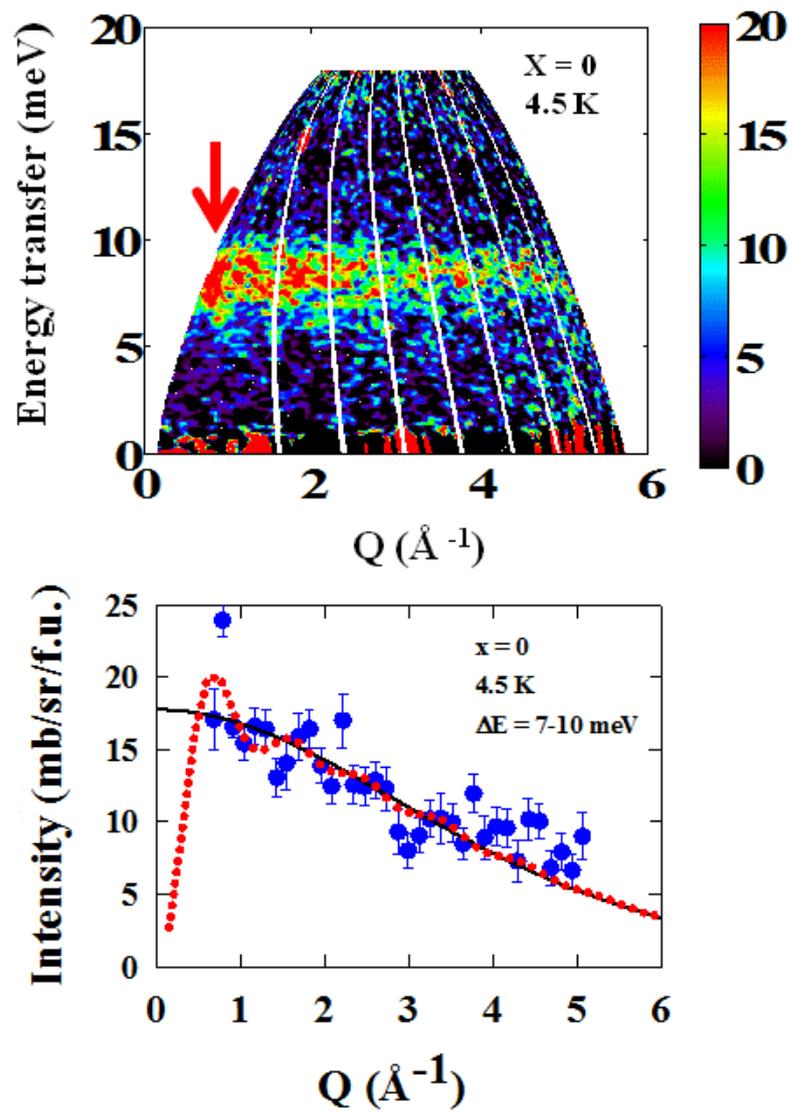

Fig.1A Adroja et al (in Appendix)

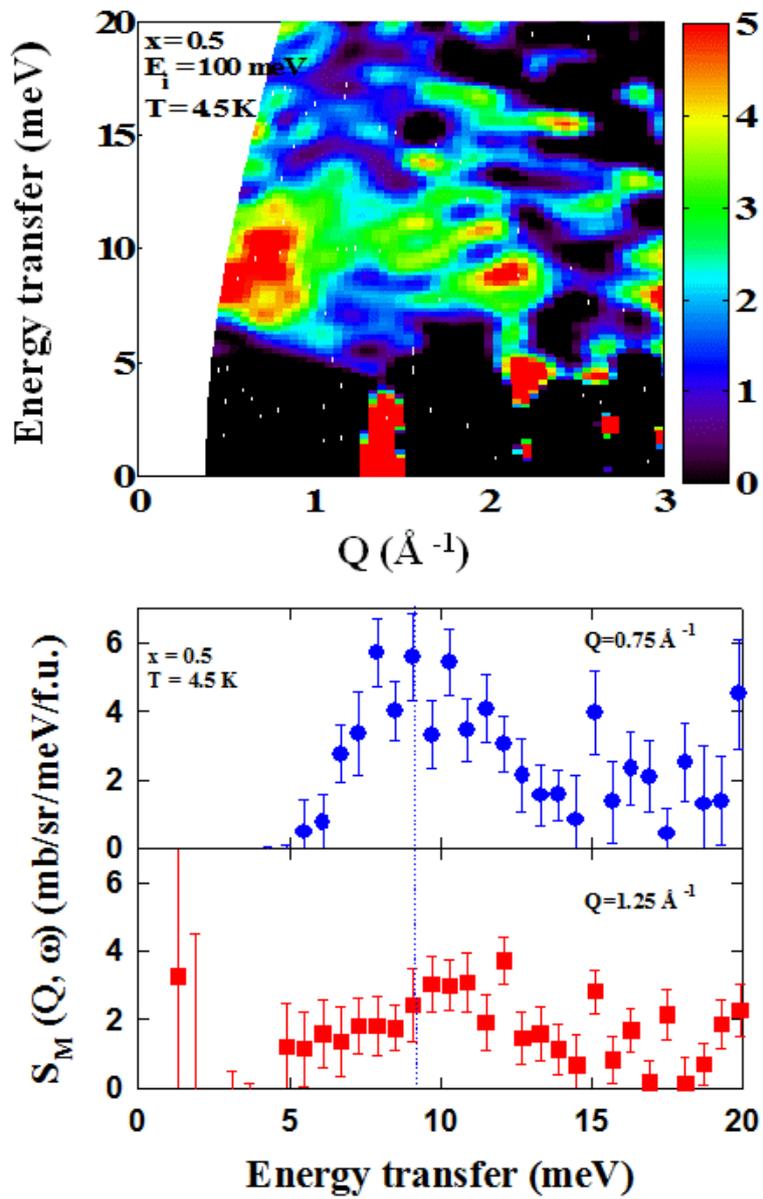

Fig. 2A Adroja et al (in Appendix)